\documentclass[letterpaper,11pt]{article}
\pdfoutput=1

\usepackage{jheppub}
\usepackage{multirow}
\usepackage{subfig}
\usepackage{xspace}
\usepackage{xcolor}
\usepackage[countmax]{subfloat}
\usepackage{amsmath}
\usepackage{mathtools}
\usepackage{graphicx}
\usepackage{slashed}
\usepackage{tikz-feynman,contour}
\usepackage{tikz,pgf}
\usepackage{braket}
\usepackage{amsfonts}
\usepackage{amsthm,scrextend,bbold}
\newcommand{\as}{\alpha_\text{s}}
\newcommand{\cf}{C_{\text{F}}}
\newcommand{\ca}{C_{\text{A}}}

\newcommand{\order}[1]{{\cal O}\left(#1\right)}

\DeclareMathOperator{\De}{d}
\newcommand{\de}{\De\!}
\newcommand{\xf}{x}

\newcommand{\xt}{\tau}

\newcommand{\xb}{x_{\text{B}}}

\newcommand{\gs}{\gamma_\text{soft}}
\newcommand{\gszero}{\gamma_\text{soft}^{(0)}}
\newcommand{\gsone}{\gamma_\text{soft}^{(1)}}
\newcommand{\cone}{C^{(1)}}

\usepackage{color}
\definecolor{darkblue}{rgb}{0,0,0.5}
\definecolor{darkgreen}{rgb}{0,0.5,0}
\definecolor{darkorange}{rgb}{0.8,0.3,0}

\title{
Soft logarithms in processes with heavy quarks
}

\author[1]{Daniele Gaggero,}
\author[1]{Andrea Ghira,}
\author[1]{Simone Marzani,}
\author[1]{and Giovanni Ridolfi}

\affiliation[1]{Dipartimento di Fisica, Universit\`a di Genova and INFN, Sezione di Genova,Via Dodecaneso 33, 16146, Italy}
\emailAdd{daniele.gaggero@ge.infn.it}
\emailAdd{andrea.ghira@ge.infn.it}
\emailAdd{simone.marzani@ge.infn.it}
\emailAdd{giovanni.ridolfi@ge.infn.it}

\abstract{Observables involving heavy quarks can be computed in perturbative QCD in two different approximation schemes: either the quark mass dependence is fully retained, or it is retained only where needed to regulate the collinear singularity. The two schemes have different advantages and drawbacks. In particular, it is known that the structure of large logarithms arising from soft emissions is different in the two approaches. We investigate the origin of this difference in some detail, focussing on a few specific processes. We show that it is related to the non-commutativity of the small-mass and soft-emission limits. Finally, we perform the resummation of soft-emission logarithms to next-to-leading accuracy in the case of Higgs decay into a $b\bar b$ pair, in the scheme in which the quark mass dependence is fully accounted for.}

\begin{document}
\maketitle

\section{Introduction}\label{sec:intro}

Quarks appear in the Quantum Chromo-Dynamics (QCD) lagrangian in different species, named flavours. From the point of view of strong interactions, different flavours are distinguished purely on the basis of the value of their masses. 
It is therefore natural to classify quark flavours according to their masses, compared to $\Lambda_{\rm QCD}$. 
The masses of up, down and strange quarks, relevant for ordinary matter, are much smaller than $\Lambda_{\rm QCD}$, and can be taken to be zero for most applications in high-energy physics.
At the opposite end of the spectrum lies the top quark, with a mass of approximately 173 GeV. This value is so large that the top quark lifetime is smaller than the typical time scale of hadron formation; for this reason, the physics of processes involving top quarks requires a dedicated treatment.
Bottom ($b$) and charm ($c$) quarks are heavy, according to the definition given above: their masses are approximately 4 GeV and 1.3 GeV respectively. Especially in the case of $b$ quarks, these values are close to the perturbative domain of QCD. 
Furthermore, from the experimental viewpoint, the lifetime of $B$ hadrons is long enough so that their decays happen away from the interaction point. Dedicated $b$-tagging techniques~\cite{ATLAS:2019bwq, CMS:2017wtu} that exploit this property to identify $B$ hadrons, or $b$ jets, are widely used in collider experiments, see e.g.~\cite{ATLAS:2020FCP,CMS:2018nsn,ATLAS:2018mme,CMS:2018uxb,ATLAS:2020juj,ATLAS:2020aln,CMS:2018fks,LHCb:2021trn,ATLAS:2020xzu, CMS:2020pyk}.
The combinations of these facts allow us to compute theoretical predictions in perturbative QCD and compare them with particle-level measurements, with confidence that non-perturbative contributions, mostly due to the hadronisation processes, are genuine power-corrections. 

Processes involving $b$ quarks are especially relevant in high energy physics, for a number of reasons. 
In the context of LHC phenomenology, the Higgs boson decays primarily into $b$ quarks and, although this decay mode is challenging because of its large background, it plays a central role in studies of electro-weak symmetry breaking. Furthermore, many aspects of so-called flavour-physics can be scrutinised in the $b$ sector and any definite statement about the presence of intrinsic heavy-flavour component (mostly charm quarks) in the proton requires precision calculations of perturbative cross-section involving heavy quarks.  
It is also worth mentioning that, although strong interactions are flavour-blind, i.e.\ gluons couple to quarks irrespectively of their mass, quark masses do affect emergent phenomena, such as jet formation and their substructure. 
In this context, the most famous effect is the so-called dead-cone, i.e.\ the fact that colour radiation around heavy quarks is suppressed~\cite{Dokshitzer:1991fd}. 
Dedicated phenomenological strategies have been designed to expose and study this effect, e.g.~\cite{Llorente:2014bha,Maltoni:2016ays,Cunqueiro:2018jbh}, which has been recently measured by the ALICE collaboration at the LHC~\cite{ALICE:2021aqk}. Furthermore, the possibility of exploiting the imprint that quark masses leave on colour correlations has been recently investigated in the context of $b$-tagging~\cite{Fedkevych:2022mid}.
Moreover, the fact that the bottom mass (and to some extent the charm mass) are in the perturbative regime, may allow us to exploit them as a probe of the so-called hadronisation process, i.e.\ the mechanism that binds quarks and gluons into colour-neutral hadrons. 
Finally, from a theoretical point of view, QCD factorisation for heavy quarks has been proven for processes with one hadron in the initial state~\cite{Collins:1998rz} but it is known that problems arise when considering hadron-hadron collisions at two loops and beyond (see~\cite{Caola:2020xup} and references therein.)

As already mentioned, heavy-flavour production cross-sections can be calculated in perturbative QCD because the mass of the $b$ and $c$ quarks sets the value of the coupling in the perturbative region and regulates collinear singularities.
Two main strategies to perform these calculations are usually employed. In  the so-called \emph{massive scheme}, the final-state heavy quarks are considered as real, on-shell particles. 
The main advantage of the massive scheme is that the kinematics of the heavy flavour is treated correctly, because the full mass-dependence is retained. 
However, perturbative calculations with massive particles are difficult at high orders. For instance, in proton collisions, heavy flavour production is known at next-to-next-to-leading order (NNLO)~\cite{Czakon:2013goa,Catani:2019iny,Catani:2020kkl}. 
The fixed-order precision can be improved by various types of all-order calculations, e.g.\ threshold resummation~\cite{Cacciari:2011hy}, high-energy resummation~\cite{Catani:1990eg,Ball:2001pq} or even transverse momentum resummation~\cite{Catani:2014qha}.
The inclusion of heavy-quark effects in general-purpose Monte Carlo parton showers is also an area of active research, see e.g.~\cite{Norrbin:2000uu,Bahr:2008pv,Krauss:2017wmx}.

The range of energies probed by collider experiments is typically much larger than the heavy-flavour mass, making heavy-flavour production a multi-scale problem. Theoretical predictions for these processes, even for inclusive observables, are plagued by logarithms of $q^2/m^2$, where $q^2$ the square of the hard scale, that can spoil the behaviour of the perturbative expansion. 
Therefore an alternative calculational framework is often employed. This approach exploits \emph{fragmentation functions} to resum these logarithmic corrections to all orders. This is possible because these logarithmic corrections are related to collinear dynamics, which would give rise to divergencies in a massless theory. It follows that, up to corrections $\order{m^2/q^2}$, heavy-flavour production cross-sections obey a factorisation theorem and can be written as the convolution of process-dependent partonic (massless) coefficient functions with universal heavy-flavour fragmentation functions. Fragmentation functions obey DGLAP evolutions equations (with time-like splitting functions), which allow one to resum the large logarithmic corrections we are discussing, in complete analogy to the initial-state collinear factorisation theorem. 
Note that in this framework we essentially treat the heavy flavour as a massless parton and its mass $m$ only acts a regulator of collinear singularities.
For a recent review of theoretical and phenomenological results in heavy flavour fragmentation, see~\cite{Corcella:2022zna} and references therein.

Virtues of the massive scheme and of fragmentation functions can be combined together by matching the fixed-order calculation performed in the massive scheme, with the all-order resummation of mass logarithms achieved by the fragmentation function approach~\cite{Cacciari:1998it,Cacciari:2001td,Forte:2010ta,Forte:2016sja,Bonvini:2015pxa,Pietrulewicz:2017gxc,Ridolfi:2019bch}.
Heavy-quark fragmentation functions have also been studied using the framework of Soft-Collinear Effective Theory (SCET), see for instance Refs~\cite{Fickinger:2016rfd,Neubert:2007je}.

There is an important difference between standard parton distribution functions (PDFs) and heavy-flavour fragmentation functions. The initial conditions for PDF are typically given at a scale at $\mu_0$ of the order of  $\Lambda_\text{QCD}$ and therefore in the non-perturbative domain, while for heavy quark fragmentation functions they are assigned at $\mu_0 \simeq m_{b,c} \gg \Lambda_\text{QCD}$. Therefore, the initial-condition is perturbative and it can be determined by matching the factorisation theorem with the massive scheme.
The initial-condition for the $b$ quark fragmentation function was determined to NLO in QCD in Ref.~\cite{Mele:1990cw,Mele:1990yq} and to NNLO in Ref.~\cite{Melnikov:2004bm,Mitov:2004du}. 
The initial condition of the evolution is, by definition, free of mass logarithms, but, as it will be discussed in detail in the following, it is affected by soft logarithms, that should be resummed to all-orders too~\cite{Cacciari:2001cw,Maltoni:2022bpy}.
It turns out that the structure of soft logarithms can significantly differ in the massive scheme and in the fragmentation function approach and this difference depends both on the considered process and on the specific observable that is computed.
The main objective of this work is to clarify the origin of this difference, which we believe has not been discussed in detail. Secondly, while explicit soft-gluon resummation formulae exists in the fragmentation function framework, resummed results that fully take into account the heavy quark mass do not appear, to the best of our knowledge, in the literature. 

The remainder of this paper is organised as follows. In Sec.~\ref{sec:fragm} we recall the basics of the massive scheme and of the fragmentation function approach, using the decay of the Higgs boson into a $b \bar b$ pair as an example. In Sec.~\ref{sec:examples} we study the interplay between the soft and the massless limit, considering not only the decay but also other processes that are related by crossing symmetry. Finally, in Sec.~\ref{sec:resum} we describe soft resummation in the massive scheme at next-to-leading logarithmic (NLL) accuracy, before drawing our conclusions in Sec.~\ref{sec:conclusions}. Details of the calculations are collected in the appendix.

\section{Massive scheme vs fragmentation functions}\label{sec:fragm}

To be definite, we focus on a specific process that allows us to highlight the different issues we would like to discuss, while maintaining the calculations as simple as possible. We consider the decay of  the Higgs boson into a $b \bar b$ pair:
\begin{equation}
h(q) \to b(p_1) + \bar{b}(p_2),
\label{decay}
\end{equation}
(momenta in brackets)
which is of clear interest for LHC phenomenology and it is also related to $Z$ decay into heavy quarks, which was extensively studied at $e^+e^-$ colliders~\cite{DELPHI:2000edu,SLD:1999cuj,ALEPH:2001pfo,OPAL:1995rqo,OPAL:1994cct,DELPHI:1992pnf}.
We are interested in the differential decay rate for this process with respect to the dimensionless ratio 
\begin{equation}\label{eq:x1}
\xf=\frac{2 p_1\cdot q}{q^2}.
\end{equation}
In the centre-of-mass frame of the Higgs boson, $x$ is fraction of the available energy (half the Higgs mass) carried by the $b$-quark. The inclusive decay rate for this process is known to N$^3$LO~\cite{Mondini:2019gid}, in the approximation in which one neglects the $b$ quark mass. 

 The calculation of the $\xf$ spectrum $\frac{\de\Gamma}{\de\xf}$ can be performed in the massive scheme, by taking into account the final-state quark mass $m$ order by order in perturbation theory. At LO the result is proportional to $\delta(1-\xf)$; the explicit NLO calculation will be performed in the next section (see also the appendices for details). Within this approach, the dependence of the $\xf$ spectrum on the heavy quark mass is taken into account exactly at each order in perturbation theory.
 On the other hand, contributions of order $\left(\as\log\frac{k_T}{m}\right)^n$, where $k_T$ is the transverse momentum of emitted gluons with respect to the $b$ quark momentum directions, are only taken into account at a finite order in perturbation theory. 
 
 Such large contributions are resummed to all orders in the fragmentation function approach. In this case, the $\xf$ spectrum is given by
 \begin{align}\label{eq:rate-frag}
 \frac{\de\Gamma}{\de\xf}&= \Gamma_0 \sum_{i} \int_x^1 \frac{\de z}{z} C_i\left(\frac{x}{z},\as, \frac{\mu^2}{q^2} \right) D_i (z,\mu^2,m^2) +\order{\frac{m^2}{q^2}},
 \end{align}
 \begin{equation}
    \Gamma_0=\frac{\sqrt{2 q^2} G_F m^2{\beta}^3 N_{\text{C}}}{8\pi}, \quad \beta=\sqrt{1-\frac{4 m^2}{q^2}}.
\end{equation}
where $G_F$ is the Fermi constant.
The sum $i$ runs over all possible partons. 
In Eq.~(\ref{eq:rate-frag}),  the functions $C_i$ are related to the rate for the production of a massless parton: a gluon, a light quark or antiquark, or a heavy quark, whose mass is however much smaller than $q^2$, and acts as a regulator of collinear divergences when higher order corrections are included. Collinear singularities appear as powers of 
$\log\frac{m^2}{q^2}$. Correspondingly, powers of $\frac{m^2}{q^2}$ are systematically neglected in the calculation of $C_i$. The factorisation scale $\mu$ is introduced to separate the collinear region, where the transverse momentum $k_T$ of emitted partons with respect to the emitting partons is much smaller than $\mu$, from the large-$k_T$ region. The contribution of the collinear region, divergent as $\frac{m^2}{q^2}\to 0$, is subtracted from the partonic coefficient functions, and absorbed in a redefinition of the fragmentation functions $D_i$,
analogously to what happens to initial-state collinear contributions. Both the coefficient functions and the fragmentation functions acquire a dependence on the arbitrary scale $\mu$. Fragmentation functions are universal, in the sense that they do not depend on the specific process we are considering, but only on the fragmenting parton. On the other hand, the coefficient functions $C_i$ obviously depend on the process.

The convolution product that appears in Eq.~(\ref{eq:rate-frag}) is turned into an ordinary product by Mellin transformation: \begin{align}\label{eq:rate-frag-N}
 \widetilde{\Gamma}(N,\xi)&= \frac{1}{\Gamma_0}\int_0^1 \de x \, x^{N-1}\,  \frac{\de\Gamma}{\de\xf}=  \sum_{i}  \widetilde{C}_i\left(N,\as, \frac{\mu^2}{q^2} \right) \widetilde{D}_i (N,\mu^2,m^2) + \order{\xi},
 \end{align}
where $ \widetilde{C}_i$ and $ \widetilde{D}_i$ indicate Mellin moments of coefficient functions and fragmentation functions, respectively, and we have defined $\xi = \frac{m^2}{|q^2|}$.\footnote{The absolute value of $q^2$ is not needed in the case of Higgs decay, where $q^2>0$; we will however keep the same definition of $\xi$ in the following,
when considering space-like configurations.}
The $\mu^2$ dependence of fragmentation functions is fixed by the requirement that
the physical cross-section does not depend on the factorisation scale. One finds
\begin{align}\label{eq:dglap}
\mu^2 \frac{\de }{\de \, \mu^2}\widetilde{D}_i (N,\mu^2,m^2) = \sum_j \gamma_{ij}\left(N,\as(\mu^2) \right) \widetilde{D}_j (N,\mu^2,m^2),
\end{align}
the DGLAP evolution equations. Here $\gamma_{ij}$ is a matrix of functions which can be extracted order by order in perturbation theory from the calculation of perturbative cross sections.
The $\gamma_{ij}$ play the role of 
anomalous dimensions; their inverse Mellin transforms are usually called the time-like splitting functions.
The solution of the differential equations~(\ref{eq:dglap}) is usually written~\cite{Cacciari:2001cw} as the product of an evolution operator $U$ and an array of initial conditions $\widetilde{D}^{(0)}$:
\begin{equation}\label{eq:dglap-sol}
\widetilde{D}_i (N,\mu^2,m^2) = \sum_j U_{ij}(N,\mu^2,\mu_0^2)  \, \widetilde{D}^{(0)}_j (N,\as, \mu_0^2,m^2),
\end{equation}
where now $U$ obeys Eq.~(\ref{eq:dglap}), with starting condition $U_{ij}(N,\mu_0^2,\mu_0^2)=\delta_{ij}$. 
As usual, by choosing $\mu_0^2=m^2$ and $\mu^2=q^2$, powers of the large logarithm of $m^2/q^2$ are resummed to all orders in the evolution factor, up to a given logarithmic accuracy, which is fixed by the order of the perturbative evaluation of the anomalous dimensions.
A detailed analysis of the $b$ quark fragmentation function in $H \to b \bar b$ has been performed, for instance, in~\cite{Corcella:2004xv}.

In order to determine the process-independent initial conditions, one chooses a process for which the massless partonic coefficient functions and the massive-scheme quantity $\widetilde{\Gamma}(N,\xi)$ are known.
Then $\widetilde{D}^{(0)}$ can be found by substituting Eq.~(\ref{eq:dglap-sol}) with $\mu=\mu_0$ into Eq.~(\ref{eq:rate-frag-N}) and by computing the $\widetilde{\Gamma}(N,\xi)$ in the $\xi \to 0$, i.e.\ $m^2\ll q^2$, limit. 

This procedure was first followed at NLO~\cite{Mele:1990cw}; the resulting order-$\as$  initial condition is seen to grow as $\as \log^2 N$ at large $N$, which corresponds, in the physical space of energy fraction, to a distribution of the form $\as \left( \frac{\log(1-z)}{1-z}\right)_+$. This is the typical double logarithmic behaviour arising from the region of soft and collinear emission by \textit{massless} partons.
These large logarithmic corrections were later studied and resummed at all orders to NLL accuracy in~\cite{Cacciari:2001cw}.

The appearance of the double-logarithmic behaviour of radiation from a \textit{massive} parton may appear as counter-intuitive: one would naively expect
that no collinear logarithms should appear, because of the finite quark mass, and therefore that the order-$\as^k$
Mellin-transformed coefficient should display a single-logarithmic behaviour $\as^k \log^k N$, eventually multiplied by mass logarithms.
We know, however, that Mellin moments of partonic coefficient functions $\widetilde{C}_i$ also display a double-logarithmic behaviour $\as^k \log^{2k}N$ to any order in perturbation theory. 
Thus, double logarithms in the fragmentation functions might in principle be compensated by analogous contributions in the coefficient functions, leaving only single-logarithms appear in the physical observable $\widetilde{\Gamma}(N,\xi)$.

We will see that this is not always the case: in some processes involving heavy quarks the double-logarithmic structure survives in the physical spectrum $\widetilde{\Gamma}(N,\xi)$, when computed in the fragmentation function scheme. The reason is that in the fragmentation function formalism the heavy quark mass is considered large with respect to $\Lambda_{\rm QCD}$, so that the initial condition for evolution can be computed perturbatively, but it is still small with respect to the hard scale of the process. Therefore, the heavy quark is treated as massless as far as radiation is concerned. In other words, the $N \to \infty$ and $\xi \to 0$ limits of $\widetilde{\Gamma}(N, \xi)$ do not always commute.

The main purpose of this paper is making more rigorous these qualitative arguments, by looking at simple examples that display the above variety of behaviours.

\section{Soft limit and massless limit}\label{sec:examples}
\begin{figure}[t]
\begin{center}
\includegraphics[width=0.3\textwidth]{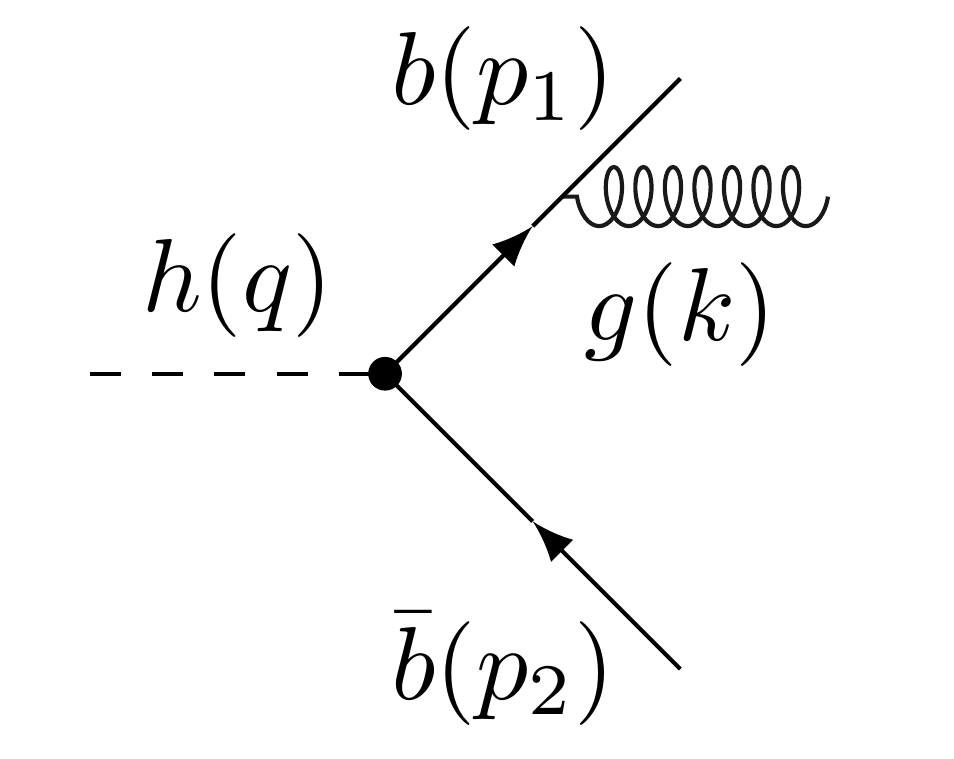}
\includegraphics[width=0.25\textwidth]{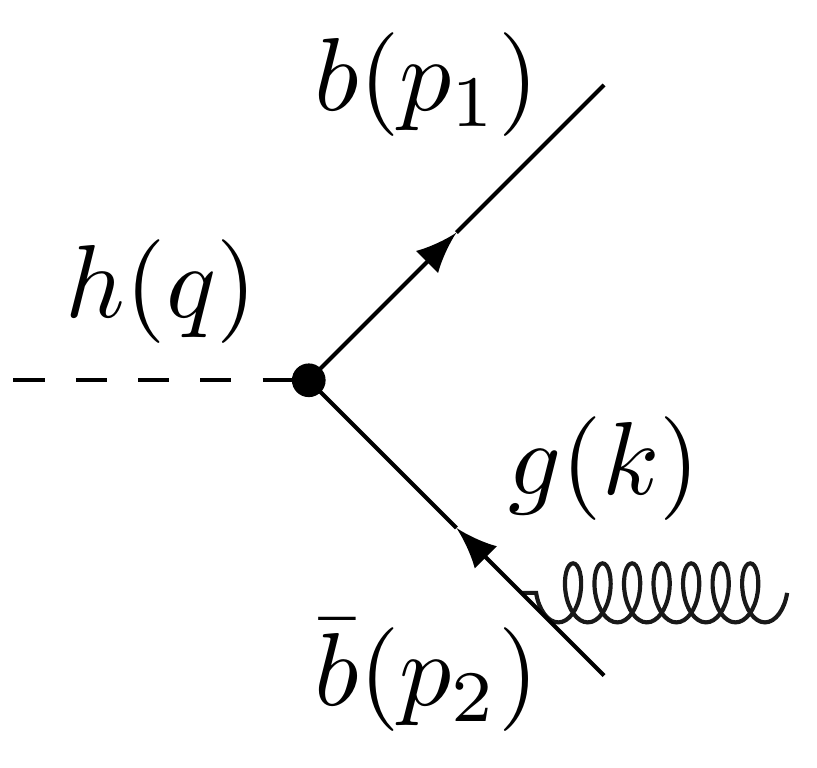}\hspace{1cm}
\caption{\label{fig:decay}
Real-emission contributions to the decay of the Higgs boson into a $b \bar b$ pair at $\order{\as}$. For the calculation in the massive scheme we keep $p_1^2=p_2^2=m^2$.}
\end{center}
\end{figure}

In this Section, we perform the explicit calculation of the differential decay rate $\frac{\de\Gamma}{\de x}$ for process (\ref{decay}) to $\order{\as}$.
Because we want to gain some understanding of the kinematics, we find it more convenient to present our discussion in $\xf$ space. We will perform the calculation in the massive scheme and then study both the small mass limit, which is necessary in order to achieve the factorisation of Eq.~(\ref{eq:rate-frag}), and the $\xf \to 1$ limit, in order to probe the soft (large $N$) region.

If we restrict our interest to the region $\xf<1$, we only have to consider the emission of a real gluon
\begin{equation}
h(q) \to b(p_1) +\bar{b}(p_2)+ g (k),
\end{equation}
with $p_1^2=p_2^2=m^2$, and $k^2=0$.
The corresponding Feynman diagrams are depicted in Fig.~\ref{fig:decay}.
The calculation of the squared invariant amplitude is straightforward.
In order to obtain the differential decay rate, it proves convenient to perform the three-body phase-space integrals in the reference frame in which $\vec{p}_2+\vec{k}=0$. We find
\begin{equation}
	\de\Phi^{(3)}(q;p_1,p_2,k)=\frac{q^2}{32(2\pi)^3} \frac{\xf(1-\xf)\beta_{\xf}}{1-\xf+\xi}\sin{\theta}\de \theta \de \xf,
\end{equation}
where $\theta$ is the angle between the quark and the antiquark 3-momenta.
After the integration over $\theta$, we obtain
\begin{align}\label{eq:GammaNLO}
\frac{1}{\Gamma_0}\frac{\de\Gamma}{\de\xf}&=\frac{\as \cf}{2 \pi} \frac{1}{\beta^{3}}
\Bigg\{\frac{\xf\beta_{\xf}}{(1-\xf)(1-\xf+\xi)}\left[ -2(1-\xf^2)+4\xi(3-4\xf+4\xi)\right] 
\nonumber \\ &
+\frac{\xf\beta_{\xf}(1-\xf)(\xf-2\xi)}{2(1-\xf+\xi)^2}+ \frac{1+\xf^2- 4 \xi (1+2 x-4 \xi)}{1-\xf} \log\frac{\xf(1+\beta_{\xf})-2\xi}{\xf(1-\beta_{\xf})-2\xi} 
\Bigg\},
\end{align}
where 
\begin{equation} \label{eq:beta-x}
\beta_{\xf}=\sqrt{1-\frac{4\xi}{\xf^2}},
\end{equation} 
which is the $b$-quark velocity in the Higgs rest frame.
Equation~(\ref{eq:GammaNLO}) simplifies considerably when all terms which are not singular in the limit $\xf\to 1$ are neglected. We find
\begin{align}    \label{Gamma_xto1}
    \frac{1}{\Gamma_0}\frac{\de\Gamma}{\de\xf}&= \frac{2 \as}{\pi}\left[  \frac{\gszero(\beta)}{1-\xf} +\mathcal{O}((1-\xf)^0) \right],
\end{align}
where have introduced, for later convenience, the (leading order) massive soft anomalous dimension
\begin{equation}\label{eq:gammasoft}
\gszero(\zeta)=  \cf \left[  \frac{1+\zeta^2}{2\zeta}\log\frac{1+\zeta}{|1-\zeta|}-1\right].
\end{equation}
Eq.~(\ref{Gamma_xto1}) tells us that in the $\xf \to 1$ limit at fixed $\xi$, i.e.\ at fixed mass of the heavy quark, the $\order{\as}$ distribution exhibits a simple pole at $\xf= 1$, which is mapped into a single logarithm in Mellin space. Furthermore, for $\xi \ll 1$, we obtain
\begin{align}    \label{Gamma_xto1_mto0}
    \frac{1}{\Gamma_0}\frac{\de\Gamma}{\de\xf}&= - \frac{\as \cf}{\pi}\left[ \frac{2\log \xi }{1-\xf} +\frac{2}{1-\xf}+\mathcal{O}((1-\xf)^0) +\mathcal{O}(\xi) \right].
\end{align}
This result matches naive expectations: a collinear logarithm multiplied by the soft singularity of the unregularised DGLAP splitting function, at leading order, $p_{qq}(x)=\frac{1+x^2}{1-x}\to\frac{2}{1-x}$.

We now take the point of view of a fragmentation function approach. In this case, we first take the $\xi \to 0$ limit of Eq.~(\ref{eq:GammaNLO}), in order to identify the mass logarithms. We find
\begin{align}    \label{Gamma_xito0}
    \frac{1}{\Gamma_0}\frac{\de\Gamma}{\de\xf}&=-\frac{\as \cf}{2\pi}\left[p_{qq}(\xf)\log\frac{\xi(1-\xf)}{\xf^2}+\frac{\xf}{1-\xf}\frac{3\xf+4}{2}+\mathcal{O}\left(\xi\right) \right]
    \nonumber \\
    &=-\frac{\as \cf}{\pi}\left[
    \frac{\log \xi}{1-\xf}+\frac{\log(1-\xf)}{1-\xf}+
    \frac{7}{4}\frac{1}{1-\xf}+\order{\xi}+ \order{(1-\xf)^0}\right],
 \end{align}
where in the second line, we have performed the $\xf\to 1$ limit  \emph{after} the $\xi \to 0$ limit.
Comparing Eqs~(\ref{Gamma_xto1_mto0}) and~(\ref{Gamma_xito0}), we realise that they present a different logarithmic structure. In the former case, we have simply the $\xf\to 1$ singularity, times the collinear logarithm, while in the latter case we have a richer structure. One of the contributions has the expected functional form $\frac{\log\xi}{1-x}$, but now a logarithmic enhancement of the form $\frac{\log (1-x)}{1-x}$ also appears, which is mapped into a double $\log N$ in Mellin space. Thus, the $\xf \to 1$ and the $\xi \to 0$ limits do not commute.

We would like to provide a physical interpretation of this mathematical fact. 
A first hint to the solution to this puzzle comes from the coefficients of $\log\xi$ in the two expressions, Eq.~(\ref{Gamma_xto1_mto0}) and Eq.~(\ref{Gamma_xito0}): there is a factor of two mismatch. It is as if the $\log(1-\xf)$ contribution in the second line of Eq.~(\ref{Gamma_xito0}) is playing the role of a collinear ($\xi$) logarithm. In the following, we shall see that this is indeed the case. 

Let us reconsider the definition of our observable $\xf$. Four-momentum conservation implies that a measurement of $\xf$ fixes the invariant mass of the antiquark-gluon system
\begin{align} \label{eq:xf-inv-mass}
m_{g\bar{b}}^2= (p_2+k)^2= (q-p_1)^2=q^2 \left(1- \xf + \xi \right).
\end{align}
Thus, a measurement of $\xf<1$ prevents the antiquark propagator in the second diagram of Fig.~\ref{fig:decay} 
from going on its mass-shell. Indeed, in the limit of small masses, the $\xf \to 1$ limit can be either associated to the emission of a soft gluon or to a gluon that is collinear to the antiquark. 
In order to analyse the actual origin of the double logarithms, we have to look instead at the quark propagator in the first diagram of Fig.~\ref{fig:decay}.
Because being differential in $\xf$ is equivalent to being differential in $m_{g\bar b}^2$, it proves convenient to perform this analysis in the rest frame of the antiquark-gluon system.
In this frame the gluon and the quark energies ($\omega_k$ and $\omega_{p_1}$, respectively) are fixed and one has to perform only one angular integration, e.g.\ over the angle $\theta$ between the gluon and the quark directions.
In this frame, the denominator of the quark propagator reads
\begin{equation}
\label{eq:quarkpropagator}
    (p_1+k)^2-m^2=2  \omega_{p_1} \omega_k (1-\beta_1 \cos{\theta}),
\end{equation}
where 
\begin{equation}\label{eq:beta1}
	\beta_1=\frac{\xf \beta_{\xf}}{\xf-2\xi}
\end{equation}
is the quark velocity in the antiquark-gluon rest frame, and $\beta_{\xf}$ is defined in Eq.~(\ref{eq:beta-x}). 
We have
\begin{align}
\label{log:divergence}
    \int^1_{-1} \frac{1}{1-\beta_1 \cos{\theta}} \de\cos{\theta}&= \frac{1}{\beta_1}\log\frac{1+\beta_1}{1-\beta_1}   \nonumber\\
    &
  = \frac{x-2 \xi}{x \beta_x}\log\frac{\xf(1+\beta_{\xf})-2\xi}{\xf(1-\beta_{\xf})-2\xi}
  = \log{\frac{\xf^2}{\xi(1-\xf)}}
    +\order{\xi},
\end{align}
where in the last step we have taken the small mass limit. 
Thus, we now understand the physical origin of the functional form of the result that one obtains if the $\xi \to 0$ limit is performed first. In this limit, collinear logarithms appear in two distinct ways: as explicit logarithm of the quark mass $m$ or as logarithms of $1-\xf$. Furthermore, we have explicitly checked, by performing the calculations with different values for the quark and the antiquark masses, that the mass appearing through the variable  $\xi$ in Eq.~(\ref{Gamma_xito0}) is the quark one, while the antiquark mass singularity is screened by the measurement of $\xf$.

The above considerations allow us to make a more general statement about the appearance of double logarithms in processes with heavy quarks, in the small mass limit. We expect this behaviour to arise if we look at a differential distribution which is directly related to the invariant mass appearing in one of the propagators. In the case we have just considered, we have indeed shown that being differential in $\xf$ is equivalent to being differential in $m_{g \bar b}^2$, which is the virtuality of one of the propagators appearing in the scattering amplitude.
Let us now consider the differential distribution in $\bar{x}=\frac{(p_1+p_2)^2}{q^2}$, i.e.\ the invariant mass of the $b \bar{b}$ system normalised to the squared Higgs mass. In this case, we do not expect any double logarithmic behaviour in the small-mass limit because no propagator has that virtuality. 
An explicit calculation gives
\begin{align} \label{eq:xbar}
\frac{1}{\Gamma_0}    \frac{\de \Gamma}{\de \bar{x}}&=\frac{\as \cf}{\pi\beta^3(1-\bar{x})}\left[2\bar{x}\beta^2\beta_{\bar{x}}+(1+\bar{x}^2-4\xi-8\xi\bar{x}+16\xi^2)\log\frac{1+\beta_{\bar{x}}}{1-\beta_{\bar{x}}}\right],
\end{align}
with $\beta_{\bar{x}}=\sqrt{1-\frac{4\xi}{\bar{x}}}$.
In the limit $\xi\to 0$ we obtain
\begin{align} \label{eq:xbar-limit}
    \frac{1}{\Gamma_0}\frac{\de\Gamma}{\de\bar{x}}&=-\frac{\as \cf}{\pi}\left(\frac{1+\bar{x}^2}{1-\bar{x}}\log\frac{\xi}{\bar{x}}+\frac{2\bar{x}}{1-\bar{x}}\right)+\mathcal{O}(\xi) \nonumber \\ 
        &= -\frac{2 \as \cf}{\pi}\frac{1+ \log \xi}{1-\bar{x}} +\mathcal{O}(\xi)+ \order{(1-\bar{x})^0},
        \end{align}
where in the second line, we have performed the limit $\bar{x}\to 1$.
Thus, as expected, in this case we only have a single logarithmic enhancement for $\bar{x}\to 1$, multiplied by a mass logarithm. Furthermore, we note that in this case the $\bar{x} \to 1$ and $\xi \to 0$ limits do commute.

\begin{figure}[t]
\begin{center}
\includegraphics[width=0.25\textwidth]{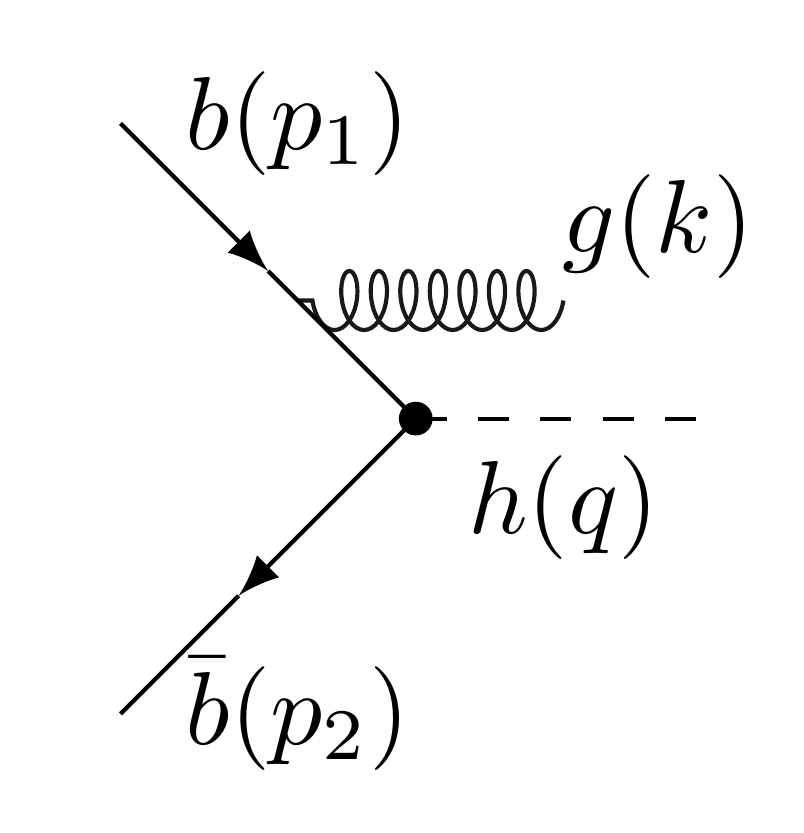}\hspace{1cm}\
\includegraphics[width=0.25\textwidth]{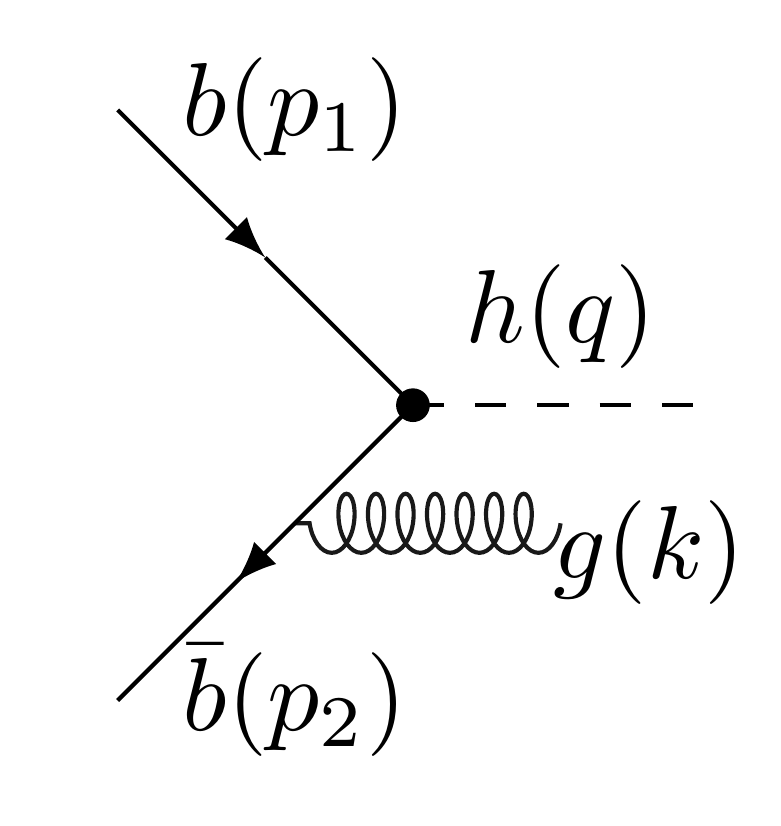}
\caption{\label{fig:production}
Real-emission contributions to the production of the Higgs boson in $b \bar b$ fusion at $\order{\as}$. For the calculation in the massive scheme we keep $p_1^2=p_2^2=m^2$.}
\end{center}
\end{figure}

We would like to test the understanding developed so far by studying other processes that involve heavy flavours:
the production of  a Higgs boson in $b \bar b$ annihilation, and the scattering of the Higgs off a $b$ quark. 
The relevant amplitudes are obtained by the same diagrams as in the case of Higgs decay, with initial and final states swapped in different ways, and therefore the corresponding squared invariant amplitudes can be easily obtained from the calculation of the previous section by crossing symmetry. 
The Higgs production cross-section has been computed to N$^3$LO accuracy in the massless 5-flavour scheme~\cite{Duhr:2019kwi}, matched to NLO in the massive 4-flavour scheme~\cite{Duhr:2020kzd}.

We start by considering  the parton-level Higgs production cross section at $\order{\as}$:
\begin{equation}
 b (p_1)+\bar{b}(p_2) \to h(q)+g(k),
\end{equation}
retaining the full mass dependence. We consider the differential distribution in the dimensionless ratio
\begin{equation}\label{eq:tau}
\xt= \frac{q^2}{(p_1+p_2)^2},
\end{equation}
which ranges between 0 and 1, and approaches 1 in the soft-radiation limit.
As in the case of Higgs decay, for $\xt$ strictly smaller than 1 we only have to evaluate real-emission diagrams, shown in Fig.~\ref{fig:production}, in four space-time dimensions. We find
\begin{equation}
\label{proddiff}
       \frac{1}{\sigma_0} \frac{\de \sigma}{\de \xt}=\ \frac{\as \cf}{\pi}\frac{1}{\beta \beta_{\xt}^2(1-\xt)}\left [-2\xt\beta^2\beta_{\xt} +\left(1+16 \xt^2 \xi^2+\xt^2 \beta^2 - 8 \xi \xt \right)\log\frac{1+\beta_{\xt}}{1-\beta_{\xt}} \right],
\end{equation}
where $\beta_{\xt}=\sqrt{1-4\xi\xt}$, and
$
\sigma_0=\frac{\sqrt{2} G_F m^2 \beta \pi N_{\text{C}}}{18 s}.
 $

We now take the limits $\xt \to 1$ and $\xi \to 0$ of this expression in different orders.
Before doing so, let us state our expectations. We are considering a process with $t$- and $u$-channel propagators, while the measurement of $\tau$ fixes the value of $s$, the invariant mass of the $b \bar b$ system. Therefore, from the analysis performed above, we expect the two limits to commute with no double-logarithmic structure appearing. 
Indeed, if we consider the limit $\xi\to 0$ of Eq.~(\ref{proddiff}) first, we find 
\begin{align}
\label{proddiff_approx_xi}
        \frac{1}{\sigma_0}\frac{\de \sigma}{\de \xt} &=-   \frac{\as C_{\text{F} }}{\pi }  \left[\frac{2\xt}{1-\xt} +p_{qq}(\tau)\log\left(\xt \xi \right) + \mathcal{O}\left(\xi\right) \right]
        \nonumber \\
        &=- \frac{2 \as C_{\text{F} }}{\pi }  \Bigg [\frac{1+\log  \xi  }{1-\xt} + \mathcal{O}\left(\xi\right) +\order{(1-\xt)^0}\Bigg ].
\end{align}
If we take the limits in the reversed order, we obtain
\begin{align}
\label{proddiff_approx_x}
        \frac{1}{\sigma_0}\frac{\de \sigma}{\de \xt} &=  \frac{2 \as  }{ \pi} \frac{\gszero (\beta)}{1-\tau} \nonumber \\
         &=- \frac{2 \as C_{\text{F} }}{\pi }\left[\frac{1+\log{\xi}}{1-\tau}+\mathcal{O}\left((1-\xt)^0\right)+\mathcal{O}(\xi)\right] .
         \end{align}
We  observe that the structure of the logarithmic singularities is the same in Eq.~(\ref{proddiff_approx_xi}) and in Eq.~(\ref{proddiff_approx_x}): no double logarithms appear and the two limits commute, in agreement with our expectations.

\begin{figure}[t]
\begin{center}
\includegraphics[width=0.25\textwidth]{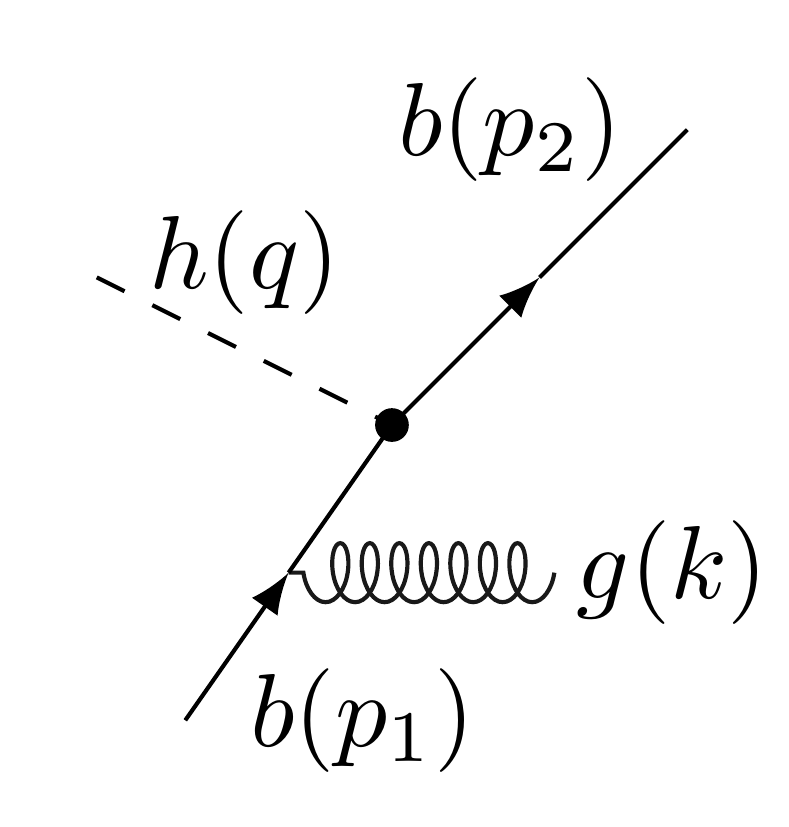}\hspace{1cm}\
\includegraphics[width=0.25\textwidth]{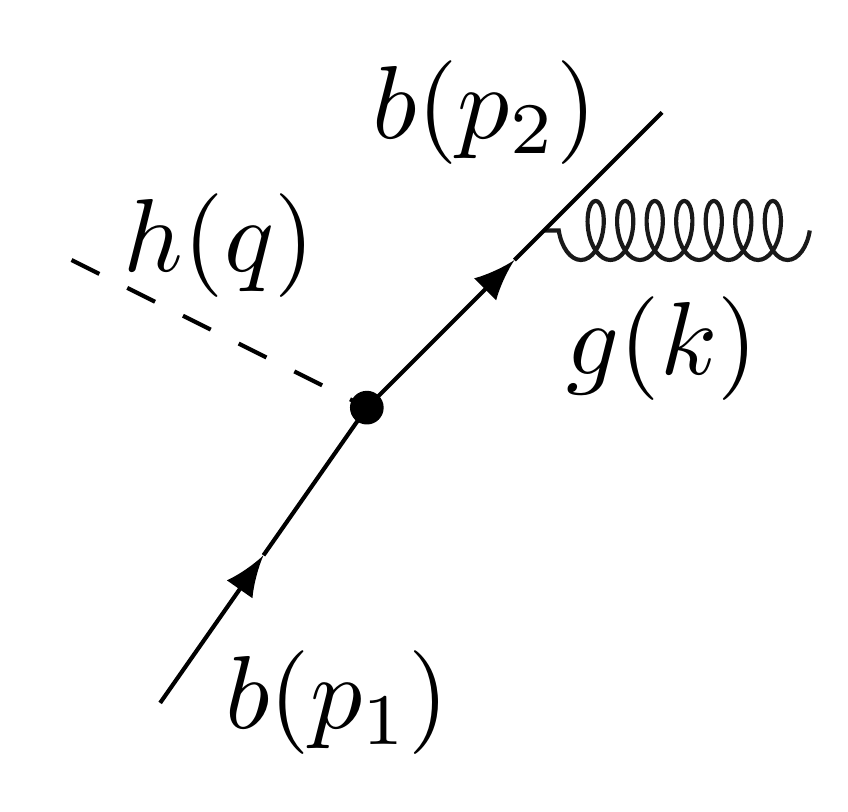}
\caption{\label{fig:scattering}
Real-emission contributions to Higgs boson scattering off a $b$ quark at $\order{\as}$. For the calculation in the massive scheme we keep $p_1^2=p_2^2=m^2$.}
\end{center}
\end{figure}

Finally, we consider the real-emission corrections to the scattering of a Higgs boson off a massive quark at NLO:
\begin{equation}
 b (p_1) +h(q)  \to b(p_2)  +g(k);
\end{equation}
the relevant diagrams are shown in Fig.~\ref{fig:scattering}.
Because now $q^2<0$, we define $Q^2=-q^2$, so that $\xi=\frac{m^2}{|q^2|}=\frac{m^2}{Q^2}>0$.
The natural choice for the scaling variable in this case is the Bjorken variable
\begin{equation} \label{eq:bjorken-x}
    \xb=\frac{Q^2}{2p_1\cdot q}=\frac{Q^2}{(p_2+k)^2-m^2+Q^2}.
\end{equation}
We obtain 
\begin{align}
\label{scattdiff}
          \frac{1}{\bar{\sigma}_0}\frac{\de\sigma}{\de \xb}&=\frac{\as\cf }{2\pi} \xb \Bigg [\frac{1+\xb^2+4\xi\xb(\xb+2)+16\xi^2\xb^2 }{(1-\xb)\eta^2_{\xb}\eta}    \log\frac{1+\eta_{\xb}+2\xi\xb}{1-\eta_{\xb}+2\xi\xb}   \nonumber \\ 
          & -
          \frac{(1-\xb)^2(4\xb+3)+\xi\xb(18\xb^2-52\xb+34)+\xi^2\xb^2(64-56\xb)+32\xi^3\xb^3}{2(1-\xb)(1-\xb+\xi \xb)^2\eta_{\xb}\eta}           
          \Bigg],
\end{align}
where $\bar{\sigma}_0=\frac{\pi \sqrt{2} G_F m^2 N_{\text{C}}\eta }{3Q^2}$ and we have introduced
\begin{equation}
\eta_{\xb}=\sqrt{1+4\xi \xb^2}, \quad \eta=\sqrt{1+4\xi}.
\end{equation}

It is clear from Eq.~(\ref{eq:bjorken-x}) that the limit $\xb \to 1$ corresponds to one of the propagators going on-shell and so, in this case, we expect the $\xb \to 1$ and $\xi \to 0$ not to commute, giving rise to double logarithmic contributions if the massless limit is taken before the soft one.
Let us check this explicitly. If take the $\xi\to 0$ first, we obtain: 
\begin{align}
\label{scattdiff_approx_xi} 
        \frac{1}{\bar{\sigma}_0}  \frac{\de\sigma}{\de \xb}  &=- \frac{\as C_{\text{F}} }{2\pi} \xb \left[ p_{qq}(\xb)\log\left(\xi \xb(1-\xb)\right)+\frac{4\xb+3}{2(1-\xb)} + \mathcal{O}(\xi) \right]
        \nonumber \\
        &=- \frac{\as C_{\text{F}} }{\pi}  \left[ 
        \frac{ \log\xi }{1-\xb}
        +\frac{\log(1-\xb)}{1-\xb}
        +\frac{7}{4}\frac{1}{1-\xb}
        + \mathcal{O}(\xi)+ \order{(1-\xb)^0}  \right],
\end{align}
which has a double-logarithmic contribution.
On the other hand, if we perform the $\xb \to 1$ limit first, we get
\begin{align}
\label{scattdiff_approx_x}
  \frac{1}{\bar{\sigma}_0}  \frac{\de\sigma}{\de \xb}  &= \frac{2\as  }{\pi}  \left[ \frac{\gszero (\eta)}{1-\xb}  +\order{(1-\xb)^0} \right]
  \nonumber \\
  &= -\frac{2 \as C_{\text{F}} }{\pi}  \left[ \frac{1+\log \xi}{1-\xb}+\order{(1-\xb)^0} + \order{\xi}\right],
\end{align}
which has no soft double-logarithms.

\section{Soft resummation with full mass dependence}\label{sec:resum}

The analysis of the previous sections led us to a better understanding of the origin of soft double-logarithmic corrections that arise in heavy-flavour production, when the fragmentation function approach is employed. In Ref.~\cite{Cacciari:2001cw} the formalism for all-order resummation of these contributions in the initial condition for the fragmentation function evolution was presented, and explicitly applied to NLL accuracy. NNLL resummation was performed in Ref.~\cite{Aglietti:2006yf}, and recently in Ref.~\cite{Maltoni:2022bpy}.
However, it is also interesting to investigate, at least from a theoretical point of view, the all-order resummation of soft logarithms in the massive scheme. This is the subject of this Section. We will consider the Higgs decay in a $b\bar b$ pair.

The differential distribution  of the $b$ quark energy fraction $x$ is an example of process with the so-called single-particle inclusive kinematics. Soft resummation in this class of processes, even in the presence of massive external lines, was carried out a long time ago~\cite{Laenen:1998qw}.~\footnote{We thank Eric Laenen for pointing out to us this result.}
The main result of Ref.~\cite{Laenen:1998qw} is a factorisation of the single-particle inclusive cross section in terms of a process-dependent hard function, a universal soft function and one jet function for each massless external parton. In our case the resummation formula simplifies considerably, because no massless parton is involved, and therefore there are no jet functions.

The resummed result of Ref.~\cite{Laenen:1998qw}, adapted to the process we are considering, reads
\begin{align}\label{resum1}
    \widetilde{\Gamma}(N,\xi)&= C(\xi,\as) \, e^{2\int_{0}^{1} d x \frac{x^{N-1}-1}{1-x}  \gs \left(\beta, \as\left((1-x)^2q^2 \right)\right)} 
     \nonumber\\
    &= \left(1+ \frac{\as}{\pi} \cone(\xi) +\order{\as^2}\right)e^{-2  \int_{1/\bar{N}}^1 \frac{d z}{z} \left[ \frac{\as(z^2 q^2)}{\pi}\gszero(\beta)+ \left(\frac{\as(z^2 q^2)}{\pi}\right)^2\gsone(\beta)+\order{\as^3}\right]} \nonumber \\&+\mathcal{O}\left(\frac{1}{N} \right),
\end{align}
with $\bar N= N e^{\gamma_E}$. Note that we have grouped the hard and soft functions at their natural scale in the process-dependent function $C(\xi,\as)$, while large-$N$ logarithms are exponentiated.  

In the last step of Eq.~(\ref{resum1}), $\widetilde{\Gamma}(N,\xi)$ is resummed to NLL accuracy, which amounts to including the soft anomalous dimension to order $\as^2$, with two-loop running coupling, and the function $C(\xi,\as)$ expanded to order $\as$. To this logarithmic accuracy we therefore need the three coefficients $\gszero, \gsone$ and $\cone$. 
The first-order coefficient of the soft anomalous dimension $\gszero$ is given
in Eq.~(\ref{eq:gammasoft}). It is process-independent, and can also be obtained from the calculation of gluon emission in the eikonal approximation, taking into account that the emitting lines are massive:
\begin{equation}
    \gszero(\beta)= (-\cf)\lim_{\varepsilon\to 0} \varepsilon \int \frac{\de ^{3-2\varepsilon} k}{2 \pi|\vec k|} \text{Eik}(p_1,p_2,k),
\end{equation}
where $\text{Eik}(p_1,p_2,k)$ is the massive eikonal factor:
\begin{equation}
\text{Eik}(p_1,p_2,k)= \frac{p_1 \cdot p_2}{p_1\cdot k\, p_2\cdot k }-\frac{p_1^2}{2(p_1\cdot k)^2}-\frac{p_2^2}{2(p_2\cdot k)^2}.
\end{equation}
The next-to-leading order massive anomalous dimension was presented in~\cite{Kidonakis:2009ev}:
\begin{align}
\gsone&= \left\{\frac{K}{2}+\frac{\ca}{2} 
\left[-\frac{1}{3}\log^2\frac{1-\beta}{1+\beta}+\log\frac{1-\beta}{1+\beta}-\zeta_2\right] \right.
\nonumber \\ &% \hspace{15mm} 
\left.
+\frac{(1+\beta^2)}{4 \beta} \ca \left[{\rm Li}_2\left(\frac{(1-\beta)^2}
{(1+\beta)^2}\right)+\frac{1}{3}\log^2\frac{1-\beta}{1+\beta}+\zeta_2\right]\right\} \, \gszero(\beta)
\nonumber \\ & %\hspace{-18mm}
{}+\cf \ca \left\{\frac{1}{2}
+\frac{1}{2} \log\frac{1-\beta}{1+\beta}
+\frac{1}{3} \log^2\frac{1-\beta}{1+\beta}
-\frac{(1+\beta^2)^2}{8 \beta^2} \left[
-{\rm Li}_3\left(\frac{(1-\beta)^2}{(1+\beta)^2}\right)+\zeta_3\right] \right.
\nonumber \\ & \left.
{}-\frac{(1+\beta^2)}{2 \beta} \left[\log\frac{1-\beta}{1+\beta}
\log\frac{(1+\beta)^2}{4 \beta}-\frac{1}{6}\log^2\frac{1-\beta}{1+\beta}
-{\rm Li}_2\left(\frac{(1-\beta)^2}{(1+\beta)^2}\right)\right]\right\},
\end{align}
with $K=\ca \left( \frac{67}{18}-\zeta_2 \right)-\frac{5n_f}{9}$.

The coefficient $\cone$ is instead process-dependent, as it receives contributions from both the end-point of the real emission and from the virtual corrections. The calculation is presented in App.~\ref{app:C1} and the result is
\begin{align}
	\label{C1}
		\cone(\xi) &=  \frac{\cf}{2}\Bigg\{-2 \frac{\gszero(\beta)}{\cf}\left[ -2\log{\left(1-\sqrt{1-\beta^2}\right)} + \log{\frac{m^2}{q^2}} +\log\left( \frac{1-\beta^2}{4}\right)+1 \right]  -2 \nonumber \\ &+ 2L(\beta)\left(\frac{1-\beta^2}{\beta}\right)+\frac{1+\beta^2}{\beta}\Bigg[\frac{1}{2}L(\beta)\log{\left(\frac{1-\beta^2}{4}\right)}+2L(\beta)(1-\log{\beta})+2\text{Li}_2\left(\frac{1-\beta}{1+\beta}\right) \nonumber \\ & +L(\beta)^2+L(\beta)\log{\frac{1-\beta}{2}} +\frac{2}{3}\pi^2 -\frac{1}{2}\left(\text{Li}_2\left(\frac{4\beta}{(1+\beta)^2} \right)-\text{Li}_2\left(\frac{-4\beta}{(1-\beta)^2} \right)\right)\Bigg] \Bigg\}  \; ,
\end{align}
 with $L(\beta)=\log\frac{1+\beta}{1-\beta}$.

It is interesting to study the small $\xi$ behaviours of the resummation coefficients. For the soft anomalous dimension, we have 
\begin{align}\label{eq:gamma-small-xi} 
\gszero(\beta)&= - \cf \log \xi+ \order{\xi^0}, \nonumber\\
\gsone(\beta)&= -\frac{\cf K}{2} \log \xi  +\order{\xi^0}.
\end{align}
Thus, because the soft anomalous dimension appears in the exponent, Eq.~(\ref{resum1}) provides a partial resummation of mass (collinear) logarithms, in particular those with soft-enhanced coefficients, which correspond to the soft part of DGLAP anomalous dimensions. However, this  resummation is only partial, even to lowest order in $\log \xi$, because it does not include hard collinear corrections.

As mentioned in Section~\ref{sec:intro}, results computed at fixed order in the strong coupling in the massive scheme are usually combined to the all-order resummation of mass logarithms, computed with the fragmentation function approach, in order achieve a more reliable result. 
The same argument holds when the  massive calculation is supplemented by soft resummation. Therefore,  one would like to match the resummed result in the massive scheme Eq.~(\ref{resum1}) not only to a fixed-order calculation in the same framework but also to a fragmentation function calculation, ideally supplemented by soft resummation~\cite{Cacciari:2001cw}.
However, as discussed at length in this paper, the situation is problematic for the $\xf$ distribution in the Higgs decay process. The massless and soft limits do not commute and, as a result, the logarithmic structure of soft resummation in the massive scheme and in the fragmentation function approach is different: the fragmentation-function calculation has double logarithms of $N$, while the massive-scheme result has only single logarithms. 
This prevents us from straightforwardly matching the two calculations.
It would be very interesting to study the construction of a matching scheme that correctly accounts for mass effects, resummation of collinear logarithms and soft logarithms, in both schemes, and to assess its relevance for heavy flavour productions at colliders. 

Finally, we note that the non-commutativity of the limits has rather spectacular consequences also for the resummed expression in the massive scheme, Eq.~(\ref{resum1}). In particular if we consider the small $\xi$ limit of the resummation constant $\cone$, we obtain
\begin{align}\label{eq:C1-small-xi} 
 \cone(\xi)&= \cf \left( \frac{1}{2}\log^2{\xi}+\log{\xi}+\order{\xi^0} \right),
\end{align}
which is double logarithmic and, therefore, in disagreement with DGLAP evolution! 
The origin of this behaviour can be traced back to the definition of the plus distribution in the massive-scheme calculation (see App~\ref{app:C1}).
We can say that double mass logarithms in the soft limit of the massive calculation and double soft logarithms in fragmentation functions are two sides of the same coin.

\section{Conclusions and outlook}\label{sec:conclusions}

We have considered different observables in processes involving heavy quarks, with different kinematics: the decay of the Higgs boson into a $b \bar b$ pair ($H\to b \bar b$), the production of the Higgs in $b \bar b$ fusion ($b \bar b \to H$) and the scattering of a Higgs boson off a $b$ quark ($b H\to b$). 
We have computed NLO corrections to these three processes, focussing on differential distributions in dimensionless ratios that measure the departure from Born kinematics. 
 All calculations were performed retaining the full quark-mass dependence, in order to study separately the massless limit and the soft limit. 
As expected, if we consider the soft limit of the massive result, only single-logarithmic corrections arise, that can be interpreted as soft logarithms. Furthermore, if we perform the $m \to 0$ limit \emph{after} soft limit,  all three processes show the same structure of logarithmically-enhanced contributions, namely a mass logarithm times a soft singularity. 

The situation radically changes if we reverse the order of the two limits. If we perform the massless limit first, which is what is done in the fragmentation function approach, and then the soft limit, the different processes give rise to distinct structures. In the $b \bar b \to H$ case, we find again the same factorised structure just described. However, in the two other cases, the two limits do not commute, and double logarithmic contributions appear.
Although this structure was discovered many years ago~\cite{Mele:1990cw}, to the best of our knowledge, a detail discussion of its origin was not present in the literature. 
Our analysis has clarified that the origin of this behaviour can be traced back to the interplay between the specific definition of  the ratio observable and the structure of the quark propagators in the scattering amplitudes. 
 For instance, in the $H\to b \bar b$ case, the variable $\xf$, defined in Eq.~(\ref{eq:x1}), can be connected to the invariant mass of the recoiling gluon-antiquark system, and the amplitude indeed features a propagator with this virtuality. Similarly, in the $b H\to b$ case, $\xb$, Eq.~(\ref{eq:bjorken-x}), is linked to the virtuality of an $s$-channel propagator, while this is not the case for $\xt$, Eq.~(\ref{eq:tau}), in the $b \bar b \to H$ and for the variable $\bar{x}=\frac{(p_1+p_2)^2}{q^2}$ in $H\to b \bar b$.
 It would interesting to revisit these studies using a different framework, such as soft-collinear effective theories.
 Indeed, the subtle behaviour of the $\xf \to 1$ limit has been pointed out both for the perturbative contribution~\cite{Fickinger:2016rfd} and non-perturbative corrections~\cite{Neubert:2007je}.
One could further study these aspects exploiting, for instance, the rich literature on resummation for boosted top production~\cite{Ferroglia:2012ku,Ferroglia:2012uy,Ferroglia:2013awa,Czakon:2018nun}, see also~\cite{Fleming:2007qr,Fleming:2007xt,Hoang:2015vua,Bachu:2020nqn}.

Finally, in Sec.~\ref{sec:resum} we have focussed on one of the processes previously considered, namely the decay of the Higgs boson. The all-order resummation of the soft-enhanced contributions to this process can be found in the literature within the fragmentation function approach, where double logarithms appear~\cite{Cacciari:2001cw}. However, as shown in Sec.~\ref{sec:examples} only single logarithms appear if the quark mass is fully retained. We have exploited results from threshold resummation in single-particle inclusive kinematics~\cite{Laenen:1998qw} to perform the resummation in the massive scheme at NLL. 
While this procedure correctly accounts for large (single) logarithms of $N$, we have discovered that in this approach one also finds double logarithms of the mass. We have traced back the origin of these contributions, which are not compatible with the standard DGLAP picture, again to the non-commutativity of the large-$N$ and small-mass limits. 

There are interesting phenomenological studies that one can imagine to carry out in the near future. For instance, we have shown that the logarithmic structure that one obtains in the fragmentation framework and in the massive scheme are not the same. It would be interesting to see whether this leads to numerical differences at collider energies. 
Finally, in the context of heavy-quark calculations, one usually combines the massive schemes with the fragmentation one, in order to obtain better predictions. 
This is a well-defined procedure, order by order in perturbation theory. However, if we want to supplement both calculations with soft gluon resummation, which differs in the two cases, the merging of the two become far from trivial. One would like to design an all-order matching scheme that consistently takes into account both massive and fragmentation approaches. We leave this to future work. 

\acknowledgments

We thank Matteo Cacciari, Stefano Catani, Stefano Frixione, and Paolo Nason for useful discussions on this topic. 
This work was supported by the Italian Ministry of Research (MUR) under grant PRIN 20172LNEEZ. 

\appendix

\section{Calculation of the resummation constant $\cone$}\label{app:C1}

The process-dependent resummation constant $\cone$ that appears in Eq.~(\ref{resum1}) can be obtained by considering the large $N$ limit of the Mellin transform of the NLO differential decay rate in the massive scheme:
\begin{equation}
\frac{ \as}{\pi} \cone(\xi) = \lim_{N \to \infty} \left[ \widetilde{\Gamma}(N,\xi) -\left(1+ \frac{ 2 \as}{\pi}\gszero(\beta) \log \frac{1}{\bar N}\right)  \right],
 \end{equation}
where the limit of large $N$ is performed at fixed $\xi$.

In order to compute the Mellin transform $\widetilde{\Gamma}(N,\xi)$ we have to extend our calculation of the differential decay rate, in order to include the end-point contribution in $\xf=1$. Thus, we have to consider both real-emission and virtual corrections at one loop
\begin{equation}
	\label{rate_tot}
	\frac{\de \Gamma}{\de\xf}=\frac{\de \Gamma_0}{\de\xf}+\frac{\de \Gamma^{(V)}}{\de\xf}+\frac{\de \Gamma^{(R)}}{\de\xf}.
\end{equation}
Because the Born and the virtual contributions have support only for $\xf=1$, we can write them as 
\begin{align}
	\frac{\de \Gamma_0}{\de\xf}=\Gamma_0\delta(1-\xf), \nonumber \\
		\frac{\de \Gamma^{(V)}}{\de\xf}=\Gamma^{(V)}\delta(1-\xf),
\end{align}
where $\Gamma^{(V)}$ is the ultra-violet renormalised one-loop contribution. Following~\cite{Braaten:1980yq}, we have Yukawa coupling, and hence the $b$ mass, in the on-shell scheme (writing the result in terms of the running mass in $\overline{\text{MS}}$ scheme is straightforward). We employ dimensional regularisation and therefore we perform our calculations in $d=4- 2 \varepsilon$ dimensions. We obtain
\begin{equation}
	\label{decay_virtual}
	\begin{split}
		\Gamma^{(V)}&= \frac{\as \cf}{2\pi} \Gamma^{(d)}_0 \Bigg\{
		\frac{2 \gszero(\beta)}{\cf } \left(\frac{1}{\varepsilon}+\log\frac{\Bar{\mu}^2}{m^2} \right) 
		\\
		&-2+ 2L(\beta)\left(\frac{1-\beta^2}{\beta}\right)+\frac{1+\beta^2}{\beta}\Bigg(\frac{1}{2}L(\beta)\log{\left(\frac{1-\beta^2}{4}\right)}-2L(\beta)\log{\beta}+2\text{Li}_2\left(\frac{1-\beta}{1+\beta}\right)\\
		&+L(\beta)^2+L(\beta)\log{\frac{1-\beta}{2}}+\frac{2}{3}\pi^2\Bigg)\Bigg\},
	\end{split}
\end{equation}
with $\Bar{\mu}^2 = 4\pi \mu^2 e^{-\gamma_{\text{E}}}$. The superscript $d$ indicates that the Born decay rate is computed in $d$ dimensions
\begin{equation}
\Gamma^{(d)}_0= \Gamma_0\,  \frac{ \pi^\frac{5-d}{2}}{2^{d-3}\Gamma\left(\frac{d-1}{2}\right)} \left(\frac{4 \mu^2}{q^2 \beta^2} \right)^\frac{4-d}{2}.
\end{equation}

In order to explicitly cancel the remaining infra-red pole, we re-evaluate the real-emission contribution in dimensional regularisation:
\begin{equation}
	\label{decay_real_def}
	\frac{\de\Gamma^{(R)}}{\de\xf}= \frac{\as\cf}{\pi} \Gamma_0^{(d)} \frac{f_\varepsilon\left(\xf,\xi,\frac{q^2}{\mu^2}\right)}{(1-\xf)^{1+2\epsilon}}  ,
\end{equation}
with
\begin{align}
	\label{f}
		f_\varepsilon\left(\xf,\xi,\frac{q^2}{\mu^2}\right) &= \frac{\sqrt{q^2}}{16\beta^3}\frac{\sqrt{\pi}(16\beta^2\pi)^{\varepsilon} {\color{red} } }{\left(\Gamma\left(\frac{3}{2}-\varepsilon\right)\right)(1-\xf+\xi)^{2-\varepsilon}}(\xf\beta_{\xf})^{1-2\varepsilon}\left(\frac{q^2}{\mu^2}\right)^{-\varepsilon} \nonumber \\ &
		\times \Bigg\{ \frac{8(1-\xf+\xi)^2\left[4\xi(1-4\xi)-(1-\varepsilon)(\xf^2+1)+2\xf(4\xi-\varepsilon)\right]}{\xf-2\xi} \nonumber \\ & \quad \quad _2F_1\left(1,\frac{1}{2},\frac{3}{2}-\varepsilon;\frac{\xf^2-4\xi}{(\xf-2\xi)^2}\right) \nonumber \\&
		+\frac{32\xi(1-4\xi)(1-\xf+\xi)^3}{(\xf-2\xi)^2}\,  _2F_1\left(1,\frac{3}{2},\frac{3}{2}-\varepsilon;\frac{\xf^2-4\xi}{(\xf-2\xi)^2}\right)+8\xi^2(8\xf-7-4\xi) \nonumber\\&
		-4\xi(1-\xf)(5-\varepsilon-\xf(9-\varepsilon))+2(1-\xf)^2(3\xf(1-\varepsilon)+4\varepsilon) \Bigg\},
\end{align}
with $_2F_1(a,b,c;z)$ the hypergeometric function. 
In order to isolate the delta contributions we may expand Eq.~(\ref{decay_real_def}) for $\varepsilon\to 0$. We have
\begin{equation}
	\label{gamma_expanded}
	\int^1_{2\sqrt{\xi}}  \frac{\de \Gamma}{\de \xf} \de\xf=\Gamma_0^{(d)}\int^1_{2\sqrt{\xi}} \de \xf \left[ \left(1+\frac{\Gamma^{(V)}}{\Gamma^{(d)}_0}\right)\delta(1-\xf)+ \frac{\as \cf}{\pi} \frac{f_{\varepsilon}\left(\xf,\xi,\frac{q^2}{\mu^2}\right)}{(1-\xf)^{1+2\varepsilon}} \right],
\end{equation}
We can expand the last term around $\varepsilon=0$ to order $\varepsilon^0$ using identity~\footnote{Note that $f_0\left(\xf,\xi,\frac{q^2}{\mu^2}\right)=f_0(\xf,\xi) $.}
\begin{align}
		\label{f_expanded}
		\frac{f_\varepsilon\left(\xf,\xi,\frac{q^2}{\mu^2}\right)}{(1-\xf)^{1+2\varepsilon}} &=   \delta(1-\xf)\left[-\frac{f_0(1,\xi)}{2\varepsilon}+f_0(1,\xi)\log(1-2\sqrt{\xi})-\frac{1}{2}\frac{\de}{\de \varepsilon}f_\varepsilon\left(1,\xi,\frac{q^2}{\mu^2}\right)\Big|_{\varepsilon=0} \right] \nonumber \\
		& +\frac{f_0(\xf,\xi)}{(1-\xf)_+} +\mathcal{O}(\varepsilon) \; .
\end{align}
As expected,  $f_0(1,\xi)= 2 \gszero(\beta)/\cf$, ensuring the cancellation of the infrared singularity that appears in Eq.~(\ref{decay_virtual}).
Furthermore note that, $\frac{\as \cf}{\pi}\Gamma_0\frac{f_0(x,\xi)}{1-x}$ corresponds to the differential decay rate computed in Eq.~(\ref{eq:GammaNLO}).
Finally
\begin{equation}
	\label{f_expanded_der}
	\begin{split}
		-\frac{1}{2}\frac{\de}{\de \varepsilon}f_\varepsilon\left(1,\xi,\frac{q^2}{\mu^2} \right)\Big|_{\varepsilon=0} &=    \frac{1}{2} \Bigg \{-\frac{2 \gszero(\beta)}{\cf}\left[\log\left( \frac{1-\beta^2}{4}\right)+1+\log  \frac{\Bar{\mu}^2}{q^2}  \right] \\-\frac{1+\beta^2}{\beta}&\left[-2L(\beta) +\frac{1}{2}\left(\text{Li}_2\left(\frac{4\beta}{(1+\beta)^2} \right)-\text{Li}_2\left(\frac{-4\beta}{(1-\beta)^2} \right)\right)  \right] \Bigg\} \;.
	\end{split}
\end{equation}

We note that, because
\begin{equation}
\int_0^1 \de x\,x^{N-1} \left(\frac{1}{1-x} \right)_+= \log \frac{1}{\bar N}+ \order{\frac{1}{N}},
\end{equation}
the constant $\cone(\xi)$ is simply given by the coefficient of the $\delta(1-x)$ term in \eqref{f_expanded}, times $C_F$.
Collecting the results in Eqs.~\eqref{decay_virtual}, \eqref{gamma_expanded}, \eqref{f_expanded} and \eqref{f_expanded_der}, we find
\begin{equation}\label{plus-xi}
\frac{1}{\Gamma_0}\frac{\de \Gamma}{\de\xf}= \delta(1-\xf) + \frac{\as}{\pi} \left[ \frac{\cf f_0(\xf,\xi)}{(1-\xf)_+} + C_1(\xi)\,\delta(1-\xf)\right],
\end{equation}
where
\begin{align}
\cone(\xi) &=  \frac{\cf}{2}\Bigg\{-2 \frac{\gszero(\beta)}{\cf}\left[ -2\log{\left(1-\sqrt{1-\beta^2}\right)} + \log{\frac{m^2}{q^2}} +\log\left( \frac{1-\beta^2}{4}\right)+1 \right]  -2 \nonumber \\ 
&+ 2L(\beta)\left(\frac{1-\beta^2}{\beta}\right)+\frac{1+\beta^2}{\beta}\Bigg[\frac{1}{2}L(\beta)\log{\left(\frac{1-\beta^2}{4}\right)}+2L(\beta)(1-\log{\beta})+2\text{Li}_2\left(\frac{1-\beta}{1+\beta}\right) \nonumber \\ & +L(\beta)^2+L(\beta)\log{\frac{1-\beta}{2}} +\frac{2}{3}\pi^2 -\frac{1}{2}\left(\text{Li}_2\left(\frac{4\beta}{(1+\beta)^2} \right)-\text{Li}_2\left(\frac{-4\beta}{(1-\beta)^2} \right)\right)\Bigg] \Bigg\}  \; ,
\end{align}
which is the result reported in the text, Eq.~(\ref{C1}). Note that the dependence on $\mu^2$ disappears, as expected. 

Finally, we note that our result Eq.~(\ref{plus-xi}) only makes sense at finite $\xi$. Indeed, if we take $\xi \to 0$, $f_0(\xf,\xi)$ develops a $\log(1-\xf)$ contribution, which makes the result ill-defined. A well-defined expression in the limit $\xi \to 0$ limit can be obtained by rewriting Eq.~(\ref{plus-xi}) as
\begin{equation}\label{different-plus}
\frac{1}{\Gamma_0}\frac{\de \Gamma}{\de\xf}= \delta(1-\xf) + \frac{\as }{\pi} \left[\cf \left(\frac{f_0(\xf,\xi)}{1-\xf}\right)_+ + A(\xi) \,\delta(1-\xf)\right],
\end{equation}
where the constant $A(\xi)$ can be determined, for instance, by integrating Eq.~(\ref{different-plus}), and comparing the result to the known NLO decay rate~\cite{Braaten:1980yq,Drees:1990dq}. Interestingly, in the $\xi \to 0$ limit, the coefficient of the delta function in Eq.~(\ref{different-plus}) exhibits a single logarithm, in accordance with DGLAP evolution:
\begin{equation}
A(\xi)=\cf \frac{3}{2} \log \xi +\order{\xi^0},
\end{equation}
which is in contrast with the spurious double-logarithmic behaviour of $\cone$, see Eq.~(\ref{eq:C1-small-xi}).

\bibliography{references}

\providecommand{\href}[2]{#2}\begingroup\raggedright\begin{thebibliography}{10}

\bibitem{ATLAS:2019bwq}
{\bf ATLAS} Collaboration, G.~Aad et~al., {\it {ATLAS b-jet identification
  performance and efficiency measurement with $t{\bar{t}}$ events in pp
  collisions at $\sqrt{s}=13$ TeV}},  {\em Eur. Phys. J. C} {\bf 79} (2019),
  no.~11 970, [\href{http://arxiv.org/abs/1907.05120}{{\tt arXiv:1907.05120}}].

\bibitem{CMS:2017wtu}
{\bf CMS} Collaboration, A.~M. Sirunyan et~al., {\it {Identification of
  heavy-flavour jets with the CMS detector in pp collisions at 13 TeV}},  {\em
  JINST} {\bf 13} (2018), no.~05 P05011,
  [\href{http://arxiv.org/abs/1712.07158}{{\tt arXiv:1712.07158}}].

\bibitem{ATLAS:2020FCP}
{\bf ATLAS} Collaboration, G.~Aad et~al., {\it {Measurements of $WH$ and $ZH$
  production in the $H \rightarrow b\bar{b}$ decay channel in $pp$ collisions
  at 13 TeV with the ATLAS detector}},  {\em Eur. Phys. J. C} {\bf 81} (2021),
  no.~2 178, [\href{http://arxiv.org/abs/2007.02873}{{\tt arXiv:2007.02873}}].

\bibitem{CMS:2018nsn}
{\bf CMS} Collaboration, A.~M. Sirunyan et~al., {\it {Observation of Higgs
  boson decay to bottom quarks}},  {\em Phys. Rev. Lett.} {\bf 121} (2018),
  no.~12 121801, [\href{http://arxiv.org/abs/1808.08242}{{\tt
  arXiv:1808.08242}}].

\bibitem{ATLAS:2018mme}
{\bf ATLAS} Collaboration, M.~Aaboud et~al., {\it {Observation of Higgs boson
  production in association with a top quark pair at the LHC with the ATLAS
  detector}},  {\em Phys. Lett. B} {\bf 784} (2018) 173--191,
  [\href{http://arxiv.org/abs/1806.00425}{{\tt arXiv:1806.00425}}].

\bibitem{CMS:2018uxb}
{\bf CMS} Collaboration, A.~M. Sirunyan et~al., {\it {Observation of
  $\mathrm{t\overline{t}}$H production}},  {\em Phys. Rev. Lett.} {\bf 120}
  (2018), no.~23 231801, [\href{http://arxiv.org/abs/1804.02610}{{\tt
  arXiv:1804.02610}}].

\bibitem{ATLAS:2020juj}
{\bf ATLAS} Collaboration, G.~Aad et~al., {\it {Measurements of the production
  cross-section for a $Z$ boson in association with $b$-jets in proton-proton
  collisions at $\sqrt{s} = 13$ TeV with the ATLAS detector}},  {\em JHEP} {\bf
  07} (2020) 044, [\href{http://arxiv.org/abs/2003.11960}{{\tt
  arXiv:2003.11960}}].

\bibitem{ATLAS:2020aln}
{\bf ATLAS} Collaboration, G.~Aad et~al., {\it {Measurement of the $t\bar{t}$
  production cross-section in the lepton+jets channel at $\sqrt{s}=13$ TeV with
  the ATLAS experiment}},  {\em Phys. Lett. B} {\bf 810} (2020) 135797,
  [\href{http://arxiv.org/abs/2006.13076}{{\tt arXiv:2006.13076}}].

\bibitem{CMS:2018fks}
{\bf CMS} Collaboration, A.~M. Sirunyan et~al., {\it {Measurement of the
  $\mathrm{t}\overline{\mathrm{t}}$ production cross section, the top quark
  mass, and the strong coupling constant using dilepton events in pp collisions
  at $\sqrt{s} =$ 13 TeV}},  {\em Eur. Phys. J. C} {\bf 79} (2019), no.~5 368,
  [\href{http://arxiv.org/abs/1812.10505}{{\tt arXiv:1812.10505}}].

\bibitem{LHCb:2021trn}
{\bf LHCb} Collaboration, R.~Aaij et~al., {\it {Test of lepton universality in
  beauty-quark decays}},  \href{http://arxiv.org/abs/2103.11769}{{\tt
  arXiv:2103.11769}}.

\bibitem{ATLAS:2020xzu}
{\bf ATLAS} Collaboration, G.~Aad et~al., {\it {Search for new phenomena with
  top quark pairs in final states with one lepton, jets, and missing transverse
  momentum in $pp$ collisions at $ \sqrt{s} $ = 13 TeV with the ATLAS
  detector}},  {\em JHEP} {\bf 04} (2021) 174,
  [\href{http://arxiv.org/abs/2012.03799}{{\tt arXiv:2012.03799}}].

\bibitem{CMS:2020pyk}
{\bf CMS} Collaboration, A.~M. Sirunyan et~al., {\it {Search for top squark
  pair production using dilepton final states in ${\text {p}}{\text {p}}$
  collision data collected at $\sqrt{s}=13\,\text {TeV} $}},  {\em Eur. Phys.
  J. C} {\bf 81} (2021), no.~1 3, [\href{http://arxiv.org/abs/2008.05936}{{\tt
  arXiv:2008.05936}}].

\bibitem{Dokshitzer:1991fd}
Y.~L. Dokshitzer, V.~A. Khoze, and S.~I. Troian, {\it {On specific QCD
  properties of heavy quark fragmentation ('dead cone')}},  {\em J. Phys. G}
  {\bf 17} (1991) 1602--1604.

\bibitem{Llorente:2014bha}
J.~Llorente and J.~Cantero, {\it {Determination of the $b$-quark mass $m_b$
  from the angular screening effects in the ATLAS $b$-jet shape data}},  {\em
  Nucl. Phys. B} {\bf 889} (2014) 401--418,
  [\href{http://arxiv.org/abs/1407.8001}{{\tt arXiv:1407.8001}}].

\bibitem{Maltoni:2016ays}
F.~Maltoni, M.~Selvaggi, and J.~Thaler, {\it {Exposing the dead cone effect
  with jet substructure techniques}},  {\em Phys. Rev. D} {\bf 94} (2016),
  no.~5 054015, [\href{http://arxiv.org/abs/1606.03449}{{\tt
  arXiv:1606.03449}}].

\bibitem{Cunqueiro:2018jbh}
L.~Cunqueiro and M.~P\l{}osko\'n, {\it {Searching for the dead cone effects
  with iterative declustering of heavy-flavor jets}},  {\em Phys. Rev. D} {\bf
  99} (2019), no.~7 074027, [\href{http://arxiv.org/abs/1812.00102}{{\tt
  arXiv:1812.00102}}].

\bibitem{ALICE:2021aqk}
{\bf ALICE} Collaboration, S.~Acharya et~al., {\it {Direct observation of the
  dead-cone effect in QCD}},  \href{http://arxiv.org/abs/2106.05713}{{\tt
  arXiv:2106.05713}}.

\bibitem{Fedkevych:2022mid}
O.~Fedkevych, C.~K. Khosa, S.~Marzani, and F.~Sforza, {\it {Identification of
  b-jets using QCD-inspired observables}},
  \href{http://arxiv.org/abs/2202.05082}{{\tt arXiv:2202.05082}}.

\bibitem{Collins:1998rz}
J.~C. Collins, {\it {Hard scattering factorization with heavy quarks: A General
  treatment}},  {\em Phys. Rev. D} {\bf 58} (1998) 094002,
  [\href{http://arxiv.org/abs/hep-ph/9806259}{{\tt hep-ph/9806259}}].

\bibitem{Caola:2020xup}
F.~Caola, K.~Melnikov, D.~Napoletano, and L.~Tancredi, {\it {Noncancellation of
  infrared singularities in collisions of massive quarks}},  {\em Phys. Rev. D}
  {\bf 103} (2021), no.~5 054013, [\href{http://arxiv.org/abs/2011.04701}{{\tt
  arXiv:2011.04701}}].

\bibitem{Czakon:2013goa}
M.~Czakon, P.~Fiedler, and A.~Mitov, {\it {Total Top-Quark Pair-Production
  Cross Section at Hadron Colliders Through $O(\alpha^4_S)$}},  {\em Phys. Rev.
  Lett.} {\bf 110} (2013) 252004, [\href{http://arxiv.org/abs/1303.6254}{{\tt
  arXiv:1303.6254}}].

\bibitem{Catani:2019iny}
S.~Catani, S.~Devoto, M.~Grazzini, S.~Kallweit, J.~Mazzitelli, and H.~Sargsyan,
  {\it {Top-quark pair hadroproduction at next-to-next-to-leading order in
  QCD}},  {\em Phys. Rev. D} {\bf 99} (2019), no.~5 051501,
  [\href{http://arxiv.org/abs/1901.04005}{{\tt arXiv:1901.04005}}].

\bibitem{Catani:2020kkl}
S.~Catani, S.~Devoto, M.~Grazzini, S.~Kallweit, and J.~Mazzitelli, {\it
  {Bottom-quark production at hadron colliders: fully differential predictions
  in NNLO QCD}},  {\em JHEP} {\bf 03} (2021) 029,
  [\href{http://arxiv.org/abs/2010.11906}{{\tt arXiv:2010.11906}}].

\bibitem{Cacciari:2011hy}
M.~Cacciari, M.~Czakon, M.~Mangano, A.~Mitov, and P.~Nason, {\it {Top-pair
  production at hadron colliders with next-to-next-to-leading logarithmic
  soft-gluon resummation}},  {\em Phys. Lett. B} {\bf 710} (2012) 612--622,
  [\href{http://arxiv.org/abs/1111.5869}{{\tt arXiv:1111.5869}}].

\bibitem{Catani:1990eg}
S.~Catani, M.~Ciafaloni, and F.~Hautmann, {\it {High-energy factorization and
  small x heavy flavor production}},  {\em Nucl. Phys. B} {\bf 366} (1991)
  135--188.

\bibitem{Ball:2001pq}
R.~D. Ball and R.~K. Ellis, {\it {Heavy quark production at high-energy}},
  {\em JHEP} {\bf 05} (2001) 053,
  [\href{http://arxiv.org/abs/hep-ph/0101199}{{\tt hep-ph/0101199}}].

\bibitem{Catani:2014qha}
S.~Catani, M.~Grazzini, and A.~Torre, {\it {Transverse-momentum resummation for
  heavy-quark hadroproduction}},  {\em Nucl. Phys. B} {\bf 890} (2014)
  518--538, [\href{http://arxiv.org/abs/1408.4564}{{\tt arXiv:1408.4564}}].

\bibitem{Norrbin:2000uu}
E.~Norrbin and T.~Sjostrand, {\it {QCD radiation off heavy particles}},  {\em
  Nucl. Phys. B} {\bf 603} (2001) 297--342,
  [\href{http://arxiv.org/abs/hep-ph/0010012}{{\tt hep-ph/0010012}}].

\bibitem{Bahr:2008pv}
M.~Bahr et~al., {\it {Herwig++ Physics and Manual}},  {\em Eur. Phys. J. C}
  {\bf 58} (2008) 639--707, [\href{http://arxiv.org/abs/0803.0883}{{\tt
  arXiv:0803.0883}}].

\bibitem{Krauss:2017wmx}
F.~Krauss and D.~Napoletano, {\it {Towards a fully massive five-flavor
  scheme}},  {\em Phys. Rev. D} {\bf 98} (2018), no.~9 096002,
  [\href{http://arxiv.org/abs/1712.06832}{{\tt arXiv:1712.06832}}].

\bibitem{Corcella:2022zna}
G.~Corcella, {\it {Selected Results in Heavy Quark Fragmentation}},
  \href{http://arxiv.org/abs/2206.11518}{{\tt arXiv:2206.11518}}.

\bibitem{Cacciari:1998it}
M.~Cacciari, M.~Greco, and P.~Nason, {\it {The P(T) spectrum in heavy flavor
  hadroproduction}},  {\em JHEP} {\bf 05} (1998) 007,
  [\href{http://arxiv.org/abs/hep-ph/9803400}{{\tt hep-ph/9803400}}].

\bibitem{Cacciari:2001td}
M.~Cacciari, S.~Frixione, and P.~Nason, {\it {The p(T) spectrum in heavy flavor
  photoproduction}},  {\em JHEP} {\bf 03} (2001) 006,
  [\href{http://arxiv.org/abs/hep-ph/0102134}{{\tt hep-ph/0102134}}].

\bibitem{Forte:2010ta}
S.~Forte, E.~Laenen, P.~Nason, and J.~Rojo, {\it {Heavy quarks in
  deep-inelastic scattering}},  {\em Nucl. Phys. B} {\bf 834} (2010) 116--162,
  [\href{http://arxiv.org/abs/1001.2312}{{\tt arXiv:1001.2312}}].

\bibitem{Forte:2016sja}
S.~Forte, D.~Napoletano, and M.~Ubiali, {\it {Higgs production in bottom-quark
  fusion: matching beyond leading order}},  {\em Phys. Lett. B} {\bf 763}
  (2016) 190--196, [\href{http://arxiv.org/abs/1607.00389}{{\tt
  arXiv:1607.00389}}].

\bibitem{Bonvini:2015pxa}
M.~Bonvini, A.~S. Papanastasiou, and F.~J. Tackmann, {\it {Resummation and
  matching of b-quark mass effects in $ b\overline{b}H $ production}},  {\em
  JHEP} {\bf 11} (2015) 196, [\href{http://arxiv.org/abs/1508.03288}{{\tt
  arXiv:1508.03288}}].

\bibitem{Pietrulewicz:2017gxc}
P.~Pietrulewicz, D.~Samitz, A.~Spiering, and F.~J. Tackmann, {\it
  {Factorization and Resummation for Massive Quark Effects in Exclusive
  Drell-Yan}},  {\em JHEP} {\bf 08} (2017) 114,
  [\href{http://arxiv.org/abs/1703.09702}{{\tt arXiv:1703.09702}}].

\bibitem{Ridolfi:2019bch}
G.~Ridolfi, M.~Ubiali, and M.~Zaro, {\it {A fragmentation-based study of heavy
  quark production}},  {\em JHEP} {\bf 01} (2020) 196,
  [\href{http://arxiv.org/abs/1911.01975}{{\tt arXiv:1911.01975}}].

\bibitem{Fickinger:2016rfd}
M.~Fickinger, S.~Fleming, C.~Kim, and E.~Mereghetti, {\it {Effective field
  theory approach to heavy quark fragmentation}},  {\em JHEP} {\bf 11} (2016)
  095, [\href{http://arxiv.org/abs/1606.07737}{{\tt arXiv:1606.07737}}].

\bibitem{Neubert:2007je}
M.~Neubert, {\it {Factorization analysis for the fragmentation functions of
  hadrons containing a heavy quark}},
  \href{http://arxiv.org/abs/0706.2136}{{\tt arXiv:0706.2136}}.

\bibitem{Mele:1990cw}
B.~Mele and P.~Nason, {\it {The Fragmentation function for heavy quarks in
  QCD}},  {\em Nucl. Phys. B} {\bf 361} (1991) 626--644. [Erratum: Nucl.Phys.B
  921, 841--842 (2017)].

\bibitem{Mele:1990yq}
B.~Mele and P.~Nason, {\it {Next-to-leading QCD calculation of the heavy quark
  fragmentation function}},  {\em Phys. Lett. B} {\bf 245} (1990) 635--639.

\bibitem{Melnikov:2004bm}
K.~Melnikov and A.~Mitov, {\it {Perturbative heavy quark fragmentation function
  through $\mathcal{O}(\alpha^2_s)$}},  {\em Phys. Rev. D} {\bf 70} (2004)
  034027, [\href{http://arxiv.org/abs/hep-ph/0404143}{{\tt hep-ph/0404143}}].

\bibitem{Mitov:2004du}
A.~Mitov, {\it {Perturbative heavy quark fragmentation function through
  $\mathcal{O}(\alpha^2_s)$: Gluon initiated contribution}},  {\em Phys. Rev.
  D} {\bf 71} (2005) 054021, [\href{http://arxiv.org/abs/hep-ph/0410205}{{\tt
  hep-ph/0410205}}].

\bibitem{Cacciari:2001cw}
M.~Cacciari and S.~Catani, {\it {Soft gluon resummation for the fragmentation
  of light and heavy quarks at large x}},  {\em Nucl. Phys. B} {\bf 617} (2001)
  253--290, [\href{http://arxiv.org/abs/hep-ph/0107138}{{\tt hep-ph/0107138}}].

\bibitem{Maltoni:2022bpy}
F.~Maltoni, G.~Ridolfi, M.~Ubiali, and M.~Zaro, {\it {Resummation effects in
  the bottom-quark fragmentation function}},
  \href{http://arxiv.org/abs/2207.10038}{{\tt arXiv:2207.10038}}.

\bibitem{DELPHI:2000edu}
{\bf DELPHI} Collaboration, P.~Abreu et~al., {\it {Hadronization properties of
  b quarks compared to light quarks in e$^+$ e$^-$ $\to \overline {qq}$ from
  183-GeV to 200-GeV}},  {\em Phys. Lett. B} {\bf 479} (2000) 118--128,
  [\href{http://arxiv.org/abs/hep-ex/0103022}{{\tt hep-ex/0103022}}]. [Erratum:
  Phys.Lett.B 492, 398--398 (2000)].

\bibitem{SLD:1999cuj}
{\bf SLD} Collaboration, K.~Abe et~al., {\it {Precise measurement of the b
  quark fragmentation function in Z0 boson decays}},  {\em Phys. Rev. Lett.}
  {\bf 84} (2000) 4300--4304, [\href{http://arxiv.org/abs/hep-ex/9912058}{{\tt
  hep-ex/9912058}}].

\bibitem{ALEPH:2001pfo}
{\bf ALEPH} Collaboration, A.~Heister et~al., {\it {Study of the fragmentation
  of b quarks into B mesons at the Z peak}},  {\em Phys. Lett. B} {\bf 512}
  (2001) 30--48, [\href{http://arxiv.org/abs/hep-ex/0106051}{{\tt
  hep-ex/0106051}}].

\bibitem{OPAL:1995rqo}
{\bf OPAL} Collaboration, G.~Alexander et~al., {\it {A Study of b quark
  fragmentation into B0 and B+ mesons at LEP}},  {\em Phys. Lett. B} {\bf 364}
  (1995) 93--106.

\bibitem{OPAL:1994cct}
{\bf OPAL} Collaboration, R.~Akers et~al., {\it {A Measurement of the
  production of D*+- mesons on the Z0 resonance}},  {\em Z. Phys. C} {\bf 67}
  (1995) 27--44.

\bibitem{DELPHI:1992pnf}
{\bf DELPHI} Collaboration, P.~Abreu et~al., {\it {A Measurement of B meson
  production and lifetime using D lepton- events in Z0 decays}},  {\em Z. Phys.
  C} {\bf 57} (1993) 181--196.

\bibitem{Mondini:2019gid}
R.~Mondini, M.~Schiavi, and C.~Williams, {\it {N$^{3}$LO predictions for the
  decay of the Higgs boson to bottom quarks}},  {\em JHEP} {\bf 06} (2019) 079,
  [\href{http://arxiv.org/abs/1904.08960}{{\tt arXiv:1904.08960}}].

\bibitem{Corcella:2004xv}
G.~Corcella, {\it {Fragmentation in H ---\ensuremath{>} b anti-b processes}},
  {\em Nucl. Phys. B} {\bf 705} (2005) 363--383,
  [\href{http://arxiv.org/abs/hep-ph/0409161}{{\tt hep-ph/0409161}}]. [Erratum:
  Nucl.Phys.B 713, 609--610 (2005)].

\bibitem{Duhr:2019kwi}
C.~Duhr, F.~Dulat, and B.~Mistlberger, {\it {Higgs Boson Production in
  Bottom-Quark Fusion to Third Order in the Strong Coupling}},  {\em Phys. Rev.
  Lett.} {\bf 125} (2020), no.~5 051804,
  [\href{http://arxiv.org/abs/1904.09990}{{\tt arXiv:1904.09990}}].

\bibitem{Duhr:2020kzd}
C.~Duhr, F.~Dulat, V.~Hirschi, and B.~Mistlberger, {\it {Higgs production in
  bottom quark fusion: matching the 4- and 5-flavour schemes to third order in
  the strong coupling}},  {\em JHEP} {\bf 08} (2020), no.~08 017,
  [\href{http://arxiv.org/abs/2004.04752}{{\tt arXiv:2004.04752}}].

\bibitem{Aglietti:2006yf}
U.~Aglietti, G.~Corcella, and G.~Ferrera, {\it {Modelling non-perturbative
  corrections to bottom-quark fragmentation}},  {\em Nucl. Phys. B} {\bf 775}
  (2007) 162--201, [\href{http://arxiv.org/abs/hep-ph/0610035}{{\tt
  hep-ph/0610035}}].

\bibitem{Laenen:1998qw}
E.~Laenen, G.~Oderda, and G.~F. Sterman, {\it {Resummation of threshold
  corrections for single particle inclusive cross-sections}},  {\em Phys. Lett.
  B} {\bf 438} (1998) 173--183,
  [\href{http://arxiv.org/abs/hep-ph/9806467}{{\tt hep-ph/9806467}}].

\bibitem{Kidonakis:2009ev}
N.~Kidonakis, {\it {Two-loop soft anomalous dimensions and NNLL resummation for
  heavy quark production}},  {\em Phys. Rev. Lett.} {\bf 102} (2009) 232003,
  [\href{http://arxiv.org/abs/0903.2561}{{\tt arXiv:0903.2561}}].

\bibitem{Ferroglia:2012ku}
A.~Ferroglia, B.~D. Pecjak, and L.~L. Yang, {\it {Soft-gluon resummation for
  boosted top-quark production at hadron colliders}},  {\em Phys. Rev. D} {\bf
  86} (2012) 034010, [\href{http://arxiv.org/abs/1205.3662}{{\tt
  arXiv:1205.3662}}].

\bibitem{Ferroglia:2012uy}
A.~Ferroglia, B.~D. Pecjak, L.~L. Yang, B.~D. Pecjak, and L.~L. Yang, {\it {The
  NNLO soft function for the pair invariant mass distribution of boosted top
  quarks}},  {\em JHEP} {\bf 10} (2012) 180,
  [\href{http://arxiv.org/abs/1207.4798}{{\tt arXiv:1207.4798}}].

\bibitem{Ferroglia:2013awa}
A.~Ferroglia, S.~Marzani, B.~D. Pecjak, and L.~L. Yang, {\it {Boosted top
  production: factorization and resummation for single-particle inclusive
  distributions}},  {\em JHEP} {\bf 01} (2014) 028,
  [\href{http://arxiv.org/abs/1310.3836}{{\tt arXiv:1310.3836}}].

\bibitem{Czakon:2018nun}
M.~Czakon, A.~Ferroglia, D.~Heymes, A.~Mitov, B.~D. Pecjak, D.~J. Scott,
  X.~Wang, and L.~L. Yang, {\it {Resummation for (boosted) top-quark pair
  production at NNLO+NNLL' in QCD}},  {\em JHEP} {\bf 05} (2018) 149,
  [\href{http://arxiv.org/abs/1803.07623}{{\tt arXiv:1803.07623}}].

\bibitem{Fleming:2007qr}
S.~Fleming, A.~H. Hoang, S.~Mantry, and I.~W. Stewart, {\it {Jets from massive
  unstable particles: Top-mass determination}},  {\em Phys. Rev. D} {\bf 77}
  (2008) 074010, [\href{http://arxiv.org/abs/hep-ph/0703207}{{\tt
  hep-ph/0703207}}].

\bibitem{Fleming:2007xt}
S.~Fleming, A.~H. Hoang, S.~Mantry, and I.~W. Stewart, {\it {Top Jets in the
  Peak Region: Factorization Analysis with NLL Resummation}},  {\em Phys. Rev.
  D} {\bf 77} (2008) 114003, [\href{http://arxiv.org/abs/0711.2079}{{\tt
  arXiv:0711.2079}}].

\bibitem{Hoang:2015vua}
A.~H. Hoang, A.~Pathak, P.~Pietrulewicz, and I.~W. Stewart, {\it {Hard Matching
  for Boosted Tops at Two Loops}},  {\em JHEP} {\bf 12} (2015) 059,
  [\href{http://arxiv.org/abs/1508.04137}{{\tt arXiv:1508.04137}}].

\bibitem{Bachu:2020nqn}
B.~Bachu, A.~H. Hoang, V.~Mateu, A.~Pathak, and I.~W. Stewart, {\it {Boosted
  top quarks in the peak region with NL3L resummation}},  {\em Phys. Rev. D}
  {\bf 104} (2021), no.~1 014026, [\href{http://arxiv.org/abs/2012.12304}{{\tt
  arXiv:2012.12304}}].

\bibitem{Braaten:1980yq}
E.~Braaten and J.~P. Leveille, {\it {Higgs Boson Decay and the Running Mass}},
  {\em Phys. Rev. D} {\bf 22} (1980) 715.

\bibitem{Drees:1990dq}
M.~Drees and K.-i. Hikasa, {\it {NOTE ON QCD CORRECTIONS TO HADRONIC HIGGS
  DECAY}},  {\em Phys. Lett. B} {\bf 240} (1990) 455. [Erratum: Phys.Lett.B
  262, 497 (1991)].

\end{thebibliography}\endgroup

\end{document}